\pdfoutput=1
\documentclass[]{llncs}

\usepackage[usenames, x11names*]{xcolor}
\usepackage[T1]{fontenc}
\usepackage[utf8]{inputenc}
\usepackage[english]{babel}
\usepackage{mathtools, amsfonts, amssymb}
\usepackage[pdftex]{graphicx}
\usepackage{soul}
\usepackage{booktabs}
\usepackage{url}
\usepackage[bookmarks=false]{hyperref}
\usepackage{csvsimple}
\usepackage{csquotes}
\usepackage{subcaption}

\usepackage[style=ieee, maxnames=6, minnames=1,doi=false,isbn=false,backend=biber, sortcites,language=english, date=year]{biblatex}
\usepackage{tikz}
\usetikzlibrary{arrows.meta,backgrounds,calc,chains,matrix,positioning,shapes,shapes.geometric,	shapes.arrows, decorations.pathmorphing, decorations.pathreplacing}
\tikzset{%
	font={\footnotesize},
	vertex/.style={draw,circle,inner sep=0pt,minimum width=0.5cm,minimum height=0.5cm,font=\small, scale=0.5},
	terminal/.style={draw,regular polygon,regular polygon sides=4,inner sep=0pt,minimum width=0.5cm,minimum height=0.5cm,font=\small, scale=1.0},
	zeroterm/.style={below,inner sep=0pt,font=\small, scale=0.3}
}

\usepackage{stmaryrd}
\usepackage[binary-units=true]{siunitx}
\usepackage[textsize=tiny]{todonotes}

\usepackage{nicematrix}

\newcommand{\qop}[1]{\ensuremath{\mathit{#1}}}
\DeclarePairedDelimiterX{\abs}[1]{\lvert}{\rvert}{\ifblank{#1}{{}\cdot{}}{#1}}

\usetikzlibrary{quantikz}

\hypersetup{
	pdftitle={Efficient Construction of Functional Representations for Quantum Algorithms},
	pdfauthor={Lukas Burgholzer, Rudy Raymond, Indranil Sengupta, and Robert Wille}
}
 
\begin{document}
\title{Efficient Construction of Functional Representations for Quantum Algorithms}
\author{Lukas Burgholzer\inst{1}
	\and Rudy Raymond\inst{2}
	\and Indranil Sengupta\inst{3}
	\and Robert Wille\inst{1, 4}
	}
\institute{Institute for Integrated Circuits, Johannes Kepler University Linz, Austria\\%\\
	\and IBM Quantum, IBM Research - Tokyo, Japan\\
	\and Indian Institute of Technology Kharagpur, India\\
	\and Software Competence Center Hagenberg GmbH (SCCH), 4232 Hagenberg, Austria\\
\email{\{lukas.burgholzer,robert.wille\}@jku.at} \hfill  \email{rudyhar@jp.ibm.com}\hfill  \email{isg@iitkgp.ac.in}\\
\url{https://iic.jku.at/eda/research/quantum} }
\authorrunning{L. Burgholzer et al.}
\maketitle

\begin{abstract}
Due to the significant progress made in the implementation of quantum hardware, efficient methods and tools to design corresponding algorithms become increasingly important. Many of these tools rely on functional representations of certain building blocks or even entire quantum algorithms which, however, inherently exhibit an exponential complexity. Although several alternative representations have been proposed to cope with this complexity, the \emph{construction} of those representations remains a bottleneck. In this work, we propose solutions for \emph{efficiently constructing} representations of quantum functionality based on the idea of conducting as many operations as possible on as small as possible intermediate representations---using Decision Diagrams as a representative functional description.
Experimental evaluations show that
applying these solutions allows to construct the desired representations several factors faster than with state-of-the-art methods.
Moreover, if repeating structures (which frequently occur in quantum algorithms) are explicitly exploited, exponential improvements are possible---allowing to construct the functionality of certain algorithms within seconds, whereas the state of the art fails to construct it in an entire day.

\end{abstract}

\section{Introduction}
\label{sec:intro}
Quantum computing promises to outperform classical computers in certain applications. While the theoretical background was already developed in the previous century, it is today that actual physical devices are evolving to a point where first experiments are performed that are suggested not to be easy on a classical computer. 
However, having hardware without efficient tools to design corresponding algorithms on it 
certainly presents an unsatisfactory situation. 
Accordingly, researchers and engineers started to develop methods and tools for important tasks such as 
synthesis/compilation~\cite{niemannImprovedSynthesisClifford2018,zulehnerEfficientMethodologyMapping2019,zulehnerCompilingSUQuantum2019,itokoQuantumCircuitCompilers2019, smithQuantumLogicSynthesis2019}, 
(classical) simulation~\cite{zulehnerAdvancedSimulationQuantum2019,pednaultLeveragingSecondaryStorage2019, villalongaFlexibleHighperformanceSimulator2019}, and verification~\cite{viamontesCheckingEquivalenceQuantum2007,yamashitaFastEquivalencecheckingQuantum2010, burgholzerVerifyingResultsIBM2020a,burgholzerAdvancedEquivalenceChecking2021}---leading to elaborate design flows and tool chains as realized, e.g., by~IBM's Qiskit~\cite{aleksandrowiczQiskitOpensourceFramework2019}, Google's Cirq~\cite{CirqPythonFramework}, and Microsoft's~QDK~\cite{QuantumDevelopmentKit}.\clearpage

These tools and the corresponding design tasks, however, frequently rely on representations of certain building blocks' functionality or even the functionality of an entire quantum algorithm.
 This poses a severe challenge since quantum functionality is most generally described by matrices of exponential dimension with respect to the size of the quantum system, i.e., $2^n\times 2^n$ for a system consisting of $n$ qubits (the quantum analogue to bits).
To date, industrial tool chains like IBM's Qiskit hardly offer efficient and scaleable solutions for constructing and representing quantum functionality (as witnessed by the evaluations later in Section~\ref{sec:results}).

Fortunately,
different approaches have been proposed that try to deal with this complexity, e.g., based on arrays~\mbox{\cite{gutierrezQuantumComputerSimulation2010,guerreschiIntelQuantumSimulator2020,jonesQuESTHighPerformance2018, gheorghiuQuantumModernQuantum2018}}, tensor networks~\cite{markovSimulatingQuantumComputation2008,wangSimulationsShorAlgorithm2017, biamonteTensorNetworksNutshell2017, kissingerPyZXLargeScale2019}, and Decision Diagrams~\cite{niemannQMDDsEfficientQuantum2016, wangXQDDbasedVerificationMethod2008, zulehnerHowEfficientlyHandle2019}.
Although we may be able to represent (i.e.,~store) the overall functionality of certain building blocks or an entire quantum algorithm using these techniques, we may not be able to construct this representation in feasible time---which constitutes a severe bottleneck for many applications in the domain of quantum computing. 
This is caused by the fact that, even though individual quantum operations typically emit a sparse, tensor product structure, their composition requires subsequent \emph{matrix-matrix} multiplications---leading to a potential decrease in sparsity and/or exploitable structure.
Hence, many computations on potentially large intermediate representations have to be conducted in order to construct the overall functional representation. 

In this paper, we propose two solutions to overcome this bottleneck---using Decision Diagrams (DDs) as a representative functional description. 
First, a general solution is presented which can be applied to arbitrary functionality and is based on the idea to conduct as many operations as possible on as small as possible intermediate representations. Besides that, another solution is proposed which explicitly exploits the fact that many quantum algorithms contain repeating structures (e.g.,  Grover iterations, random walks, etc.).
In both cases, the complexity of constructing quantum functionality representations is substantially reduced---in case of the second solution even an exponential improvement is achieved.

Experimental evaluations eventually confirm the resulting benefits. 
They show that the proposed solutions allow to construct
the desired representations several factors faster than with the current state of the art. 
If additionally repeating structures are exploited, representations for quantum algorithms and building blocks can be constructed in a matter of seconds which, using the current state of the art, could not be constructed in an entire day. 
The resulting implementation is available as open source at \url{https://github.com/iic-jku/qfr}.

The rest of this paper is structured as follows: Section~\ref{sec:background} reviews the necessary basics on quantum computing and introduces the Quantum Fourier Transform, which will be used as a running example in this paper.
In Section~\ref{sec:building}, we show the importance of the considered problem and review the state of the art---illustrating the current bottleneck.
Then, Section~\ref{sec:proposed} introduces and describes the proposed solution which, afterwards, is evaluated in Section~\ref{sec:results}.
Finally, Section~\ref{sec:conclusions} concludes the paper.

\section{Background}
\label{sec:background}
In this section, we briefly review the key concepts of quantum computing as well as a typical building block for quantum algorithms which will serve as an example over the course of this paper. While the respective reviews are kept brief, we refer the interested reader to \cite{nielsenQuantumComputationQuantum2010} for a more thorough treatment on quantum computing.

In classical computing, \emph{bits} are used as the smallest computation unit---attaining values from the discrete set $\mathbb{B}=\{0,1\}$.
In the field of quantum computing, these discrete values, denoted $\ket{0}$ and $\ket{1}$ using Dirac notation, are chosen as basis elements spanning a two-dimensional complex Hilbert space~$\mathbb{H}$. 
Consequently, the state~\ket{q} of a \emph{qubit} (the quantum analogue to the bit) is described by an element of this space, i.e., by a \emph{superposition} of the basis states~$\ket{0}$ and~$\ket{1}$. More specifically, \mbox{$\ket{q} = \alpha_0 \ket{0} + \alpha_1 \ket{1}$} with $\alpha_i\in\mathbb{C}$ such that \mbox{$\abs{\alpha}^2 = \abs{\alpha_0}^2 + \abs{\alpha_1}^2 = 1$}. 

A \emph{quantum system} then consists of $n$ qubits $q_0,\dots,q_{n-1}$ described by the $2^n$-dimensional Hilbert space $\mathbb{H}\otimes\cdots\otimes\mathbb{H}$.
The state~$\ket{q}_n$ of such a system is again described by amplitudes $\alpha_i \in\mathbb{C}$, where $\ket{q}_n = \sum_{i\in\{0,1\}^n} \alpha_i \ket{i}$ with \mbox{$\abs{\alpha}^2 = 1$}.
However, the amplitudes $\alpha_i$ of a quantum system are not directly observable. Instead, performing a \emph{measurement} probabilistically collapses the qubits' state to one of the basis states $\ket{i}$ (each with probability $\abs{\alpha_i}^2$).

The state of a quantum system is manipulated through unitary linear transformations 
\mbox{$U\colon\; \mathbb{H}\otimes\cdots\otimes\mathbb{H} \rightarrow \mathbb{H}\otimes\cdots\otimes\mathbb{H}$},
which are predominantly described by their unitary \mbox{$2^n \times 2^n$} matrix representations\footnote{A complex-valued matrix $U$ is unitary if $U^\dag U = UU^\dag = \mathbb{I}$, where $U^\dag$ denotes the conjugate transpose of $U$ and $\mathbb{I}$ the identity matrix.} in the computational basis \mbox{$\{\ket{0},\dots,\ket{2^n-1}\}$}. Usually, these \emph{quantum operations} act only on a small subset of a system's qubits. Hence, their matrix representations have a sparse, tensor product structure, where the tensor product of smaller \enquote{operation matrices} with identity matrices is formed.

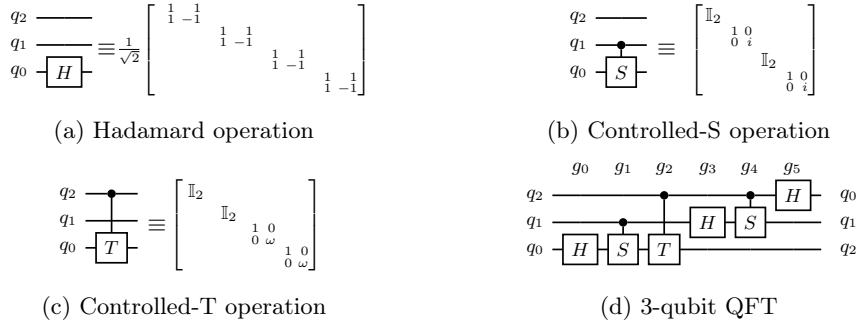
\begin{figure}[t]
	\begin{subfigure}[b]{.45\linewidth}
		\centering
		\begin{tikzpicture}
			\node[scale=0.8] (g0) {
			\begin{quantikz}[column sep=5pt, row sep={0.45cm,between origins}, ampersand replacement=\&]
			\lstick{$q_2$} \&  \qw \& \qw \\
			\lstick{$q_1$} \& \qw \& \qw\\
			\lstick{$q_0$} \&\gate{H} \& \qw
			\end{quantikz}
			};

			\node[scale=0.8] (m0) at ($(g0) + (2.6,0)$) {
			$\frac{1}{\sqrt{2}}\begin{bNiceArray}{RRRRRRRR}[small, columns-width=auto]
				1 & 1 &  &  &  &  &  &  \\
				1& -1 &  &  &  &  &  &  \\
				&  &1 & 1 &  &  &  &  \\
				&  & 1& -1 &  &  &  &  \\
				&  &  &  &1 & 1 &  &   \\
				&  &  &  & 1& -1 &  &  \\
				&  &  &  &  &  &1 & 1  \\
				&  &  &  &  &  & 1& -1  \\
				\end{bNiceArray}$
			};
			
			\node (e) at ($(g0)!0.3!(m0)$) {$\equiv$};
		\end{tikzpicture}
		\caption{Hadamard operation}
		\label{fig:h_matrix}
	\end{subfigure}\hfill
\begin{subfigure}[b]{.45\linewidth}
	\centering
		\begin{tikzpicture}
			\node[scale=0.8] (g1) {
			\begin{quantikz}[column sep=5pt, row sep={0.45cm,between origins}, ampersand replacement=\&]
			\lstick{$q_2$} \&  \qw \& \qw \\
			\lstick{$q_1$} \& \ctrl{1} \& \qw\\
			\lstick{$q_0$} \&\gate{S} \& \qw
			\end{quantikz}
			};

			\node[scale=0.8] (m1) at ($(g1) + (2.,0)$) {
			$\begin{bNiceArray}{RRRRRRRR}[small, columns-width=auto]
			\Block{2-2}{\mathbb{I}_2} &  &  &  &  &  &  &  \\
			\\
			&  &1 & 0 &  &  &  &  \\
			&  & 0& i &  &  &  &  \\
			&  &  &  &\Block{2-2}{\mathbb{I}_2} &  &  &  \\
			\\
			&  &  &  &  &  &1 & 0  \\
			&  &  &  &  &  & 0& i  \\
			\end{bNiceArray}$
		};
			
			\node (e) at ($(g1)!0.41!(m1)$) {$\equiv$};
		\end{tikzpicture}
	\caption{Controlled-S operation}
	\label{fig:s_matrix}
\end{subfigure}%

	\begin{subfigure}[b]{.45\linewidth}
	\centering
			\begin{tikzpicture}
			\node[scale=0.8] (g2) {
			\begin{quantikz}[column sep=5pt, row sep={0.45cm,between origins}, ampersand replacement=\&]
			\lstick{$q_2$} \&  \ctrl{2} \& \qw \\
			\lstick{$q_1$} \& \qw \& \qw\\
			\lstick{$q_0$} \&\gate{T} \& \qw
			\end{quantikz}
			};

			\node[scale=0.8] (m2) at ($(g1) + (2.,0)$) {
			$\begin{bNiceArray}{RRRRRRRR}[small, columns-width=auto]
			\Block{2-2}{\mathbb{I}_2} & & & & & & & \\
			 &  & & & & & &\\
			 & & \Block{2-2}{\mathbb{I}_2}& & & & &\\
			 & & & & & & &\\
			 & & & &1 & 0 & & \\
			 & & & & 0& \omega & &  \\
			 & & & & & &1 & 0  \\
			 & & & & & & 0& \omega  \\
			\end{bNiceArray}$
		};
			
			\node (e) at ($(g2)!0.41!(m2)$) {$\equiv$};
		\end{tikzpicture}
	\caption{Controlled-T operation}
	\label{fig:t_matrix}
	\end{subfigure}\hfill
	\begin{subfigure}[b]{.45\linewidth}
	\centering
	\newlength{\rbox}
	\setlength{\rbox}{0.1mm}
	\begin{tikzpicture}
			\node[scale=0.8] (g1) {
			\begin{quantikz}[column sep=5pt, row sep={0.45cm,between origins}, ampersand replacement=\&]
			\& \lstick[label style={xshift=0.3cm}]{$g_0$} \& \lstick[label style={xshift=0.3cm}]{$g_1$} \& \lstick[label style={xshift=0.3cm}]{$g_2$} \& \lstick[label style={xshift=0.3cm}]{$g_3$} \& \lstick[label style={xshift=0.28cm}]{$g_4$} \& \lstick[label style={xshift=0.28cm}]{$g_5$} \& \\
			\lstick{$q_2$} \& \qw \& \qw \&  \ctrl{2} \& \qw \& \ctrl{1} \& \gate{H} \& \qw \& \rstick{$q_0$} \\
			\lstick{$q_1$} \& \qw \& \ctrl{1} \& \qw \& \gate{H} \& \gate{S} \& \qw \& \qw\& \rstick{$q_1$}\\
			\lstick{$q_0$} \& \gate{H} \& \gate{S} \& \gate{T} \& \qw \& \qw \& \qw \& \qw\& \rstick{$q_2$}
			\end{quantikz}};
	\end{tikzpicture}
\caption{3-qubit QFT}
\label{fig:qft3}
	\end{subfigure}%
	\caption{Common quantum operations and the QFT in a 3-qubit system}
	\label{fig:operations}
\end{figure}

\begin{example}\label{ex:ops}
Consider a quantum system consisting of $n=3$ qubits. Then, Fig.~\ref{fig:h_matrix}, Fig.~\ref{fig:s_matrix}, and Fig.~\ref{fig:t_matrix} show a few common quantum operations using their $2^3 \times 2^3$ sparse matrix representations---namely the Hadamard operation as well as the controlled-phase operations~\qop{S} and~\qop{T}, where~$\omega=\exp(\frac{2 \pi i}{8}) = \sqrt{i}$.
\end{example}

A \emph{quantum algorithm} is described as a sequence of quantum operations applied to a quantum system, i.e.,~\mbox{$G=g_0,\dots,g_{m-1}$} denotes a quantum algorithm consisting of $m$ operations where each~$g_i$ is described by a unitary matrix $U_i$.
Since the composition of unitary transformations is again unitary, the functionality of a quantum algorithm may be interpreted as one unitary transformation itself. Consequently, the functionality is described by a unitary matrix~$U$ which arises from the \emph{matrix-matrix} multiplication of the individual operation matrices~$U_i$, i.e., $U=U_{m-1}\cdots U_0$.
Quantum algorithms are usually visualized as \emph{quantum circuit diagrams} where wires indicate the individual qubits and operations (also called \emph{gates}) are placed as boxes on these lines with corresponding identifiers. Time is assumed to progress from left to right. 

In the following, quantum algorithms and quantum circuit diagrams will be illustrated by means of the \emph{Quantum Fourier Transform} (QFT)~\cite{nielsenQuantumComputationQuantum2010}---a well-known 
building block in many important quantum algorithms. Its most prominent use probably is for period finding in Shor’s algorithm for integer factorization~\cite{shorPolynomialtimeAlgorithmsPrime1997} 
and many other group-theoretic problems (see Chapter 5 of~\cite{nielsenQuantumComputationQuantum2010}), at which exponential \mbox{speed-ups} over the best
classical methods are demonstrated. QFT is also used in quantum approximate counting~\cite{brassardQuantumAmplitudeAmplification2002}, which provides proven polynomial \mbox{speed-ups} over the 
best classical methods, i.e., Monte-Carlo-type estimators~\cite{montanaroQuantumSpeedupMonte2015}. Such quantum Monte-Carlo algorithms are now popular candidates to achieve quantum advantage 
with near-term quantum devices~\cite{rebentrostQuantumComputationalFinance2018}.

\begin{example}
Consider a 3-qubit system as already discussed in Example~\ref{ex:ops}.
 Then, Fig.~\ref{fig:qft3} shows the quantum circuit for the \mbox{3-qubit} Quantum Fourier Transform consisting of $m=6$ gates in total. This circuit will be used as a running example for the further discussions throughout this work.
\end{example}

\section{Representations for Quantum Algorithms}
\label{sec:building}

Working in the domain of quantum computing requires representations of certain building blocks or even entire quantum algorithms.
This is evident, e.g., for typical tasks such as:

\begin{itemize}
\item \emph{Synthesis}/\emph{Compilation}~\cite{niemannImprovedSynthesisClifford2018,zulehnerEfficientMethodologyMapping2019,itokoQuantumCircuitCompilers2019, smithQuantumLogicSynthesis2019, zulehnerCompilingSUQuantum2019}, where an entire quantum algorithm is realized in terms of elementary quantum operations supported by the addressed quantum architecture. Without a proper representation of the algorithm's functionality, no synthesis/compilation approach can work.

\item \emph{(Classical) Simulation}~\cite{zulehnerAdvancedSimulationQuantum2019,pednaultLeveragingSecondaryStorage2019, villalongaFlexibleHighperformanceSimulator2019}, where a given quantum algorithm is ``tested'' on a classical machine prior to actual execution on a quantum computer. While this can be done using consecutive matrix-vector multiplication on the elementary gates, approaches based on \emph{emulation}~\cite{steigerProjectQOpenSource2018,zulehnerMatrixVectorVsMatrixMatrix2019}, which utilize functional representations of entire building blocks, have been shown to be much more efficient---provided the emulated functionality can be constructed efficiently.

\item \emph{Verification}~\cite{viamontesCheckingEquivalenceQuantum2007,yamashitaFastEquivalencecheckingQuantum2010, burgholzerVerifyingResultsIBM2020a, burgholzerAdvancedEquivalenceChecking2021}, where, e.g., for two quantum circuits $G$ and $G^\prime$ it should be checked whether they realize the same function---also referred to as \emph{equivalence checking}. This obviously requires the construction of a functional representation for both functionalities in order to compare them. 
\end{itemize}

In all these cases, having a representation of the considered functionality is essential.
The first challenge resulting from that is that quantum functionality in general is described in terms of matrices with exponential size, i.e., for a functionality over $n$ qubits,  a matrix $U$ of size $2^n\times 2^n$ results.
In previous work, researchers already started to address this challenge, which led to different approaches exploiting certain structural elements of the considered functionality in order to reduce the exponential space complexity of its representation:

\begin{itemize}
	\item \emph{Array-based approaches} (such as proposed in~\cite{gutierrezQuantumComputerSimulation2010,guerreschiIntelQuantumSimulator2020,jonesQuESTHighPerformance2018, gheorghiuQuantumModernQuantum2018}) heavily rely on the sparsity of the involved matrices and try to distribute the workload over several cores of supercomputers, which can often be done efficiently since the \emph{matrix-multiplication} itself is inherently parallelizable. 
	\item \emph{Tensor Networks} (such as proposed in~\cite{markovSimulatingQuantumComputation2008,wangSimulationsShorAlgorithm2017, biamonteTensorNetworksNutshell2017, kissingerPyZXLargeScale2019}) capitalize on the tensor product structure inherent to quantum operations---allowing to decompose the whole matrix into many smaller parts. Their performance typically scales with the degree of entanglement of the considered functionality.

	\item \emph{Decision Diagrams} (DDs, such as proposed in~\mbox{\cite{niemannQMDDsEfficientQuantum2016, wangXQDDbasedVerificationMethod2008, zulehnerHowEfficientlyHandle2019}}) recursively split the considered functionality into equally sized sub-matrices until only complex numbers remain. By identifying redundancies in these \mbox{sub-parts} and extracting common factors, equal sub-functionality can be shared---frequently leading to a compact representation in terms of directed acyclic graphs with edge-weights. 
\end{itemize}

In the following, we will illustrate those endeavours using Decision Diagrams as a representative. However, the observations and findings discussed in this work apply to the other representations as well.

\begin{figure}[t]
\centering
	\resizebox{0.88\linewidth}{!}{                         
	\begin{tikzpicture}

\node[scale=0.7] (g0) {
			\begin{quantikz}[column sep=5pt, row sep={0.45cm,between origins}, ampersand replacement=\&]
			\lstick{$q_2$} \&  \qw \& \qw \\
			\lstick{$q_1$} \& \qw \& \qw\\
			\lstick{$q_0$} \&\gate{H} \& \qw
			\end{quantikz}
			};
	\node[right=-2mm of g0] (e0) {$\equiv$};
		
	\matrix[right=-1mm of e0, matrix of nodes,ampersand replacement=\&,every node/.style=vertex,row sep={0.4cm,between origins},column sep={0.2cm,between origins}] (dd0) {
		\node (n2) {\small$q_2$}; \\
		\node (n1) {\small$q_1$}; \\
		\node (n0) {\small$q_0$}; \\
		\node[terminal] (t){\footnotesize 1}; \\
	};
	\draw (n2.north) to ++ (0,0.15) node[left,inner sep=0pt,font=\small, scale=0.5] {$\tfrac{1}{\sqrt{2}}$};
	\draw (n2.south west) to [out=-135,in=135] (n1.north west);
	\draw (n2.south east) to [out=-45,in=45] (n1.north east);
	\draw (n2.-105) to [out=-135, in=90] ++ (-0.03,-0.03) node[zeroterm]{0};
	\draw (n2.-75) to [out=-45, in=90] ++ (0.03,-0.03) node[zeroterm]{0};
	
	\draw (n1.south west) to [out=-135,in=135] (n0.north west);
	\draw (n1.south east) to [out=-45,in=45] (n0.north east);
	\draw (n1.-105) to [out=-135, in=90] ++ (-0.03,-0.03) node[zeroterm]{0};
	\draw (n1.-75) to [out=-45, in=90] ++ (0.03,-0.03) node[zeroterm]{0};
	
	\draw (n0.south west) to [out=-135,in=135] (t.north west);
	\draw (n0.south east) node[below right,font=\small, scale=0.5]{$-1$} to [out=-45,in=45] (t.north east);
	\draw (n0.-105) to [out=-135,in=120] (t.105);
	\draw (n0.-75) to [out=-45,in=60] (t.75);

	\node[right=0.2cm of dd0,scale=0.7] (g1) {
			\begin{quantikz}[column sep=5pt, row sep={0.45cm,between origins}, ampersand replacement=\&]
			\lstick{$q_2$} \&  \qw \& \qw \\
			\lstick{$q_1$} \& \ctrl{1} \& \qw\\
			\lstick{$q_0$} \&\gate{S} \& \qw
			\end{quantikz}
			};

	\node[right=-2mm of g1] (e1) {$\equiv$};
	
	\matrix[right=-1mm of e1,ampersand replacement=\&,every node/.style=vertex,row sep={0.4cm,between origins},column sep={0.2cm,between origins}] (dd1) {
		\&\node (n2) {\small$q_2$}; \& \\
		\&\node (n1) {\small$q_1$}; \& \\
		\node (n0a) {\small$q_0$}; \& \& \node (n0b) {\small$q_0$}; \\
		\& \node[terminal] (t){\footnotesize 1}; \& \\
	};
	\draw (n2.north) to ++ (0,0.15);
	\draw (n2.south west) to [out=-135,in=135] (n1.north west);
	\draw (n2.south east) to [out=-45,in=45] (n1.north east);
	\draw (n2.-105) to [out=-135, in=90] ++ (-0.03,-0.03) node[zeroterm]{0};
	\draw (n2.-75) to [out=-45, in=90] ++ (0.03,-0.03) node[zeroterm]{0};
	
	\draw (n1.south west) to [out=-135,in=90] (n0a.north);
	\draw (n1.south east) to [out=-45,in=90] (n0b.north);
	\draw (n1.-105) to [out=-135, in=90] ++ (-0.03,-0.03) node[zeroterm]{0};
	\draw (n1.-75) to [out=-45, in=90] ++ (0.03,-0.03) node[zeroterm]{0};
	
	\draw (n0a.south west) to [out=-135,in=180] (t.west);
	\draw (n0a.south east) to [out=-45,in=90] (t.95);
	\draw (n0a.-105) to [out=-135, in=90] ++ (-0.03,-0.03) node[zeroterm]{0};
	\draw (n0a.-75) to [out=-45, in=90] ++ (0.03,-0.03) node[zeroterm]{0};
	
	\draw (n0b.south west) to [out=-135,in=90] (t.85);
	\draw (n0b.south east) node[below right,font=\small, scale=0.5] {$i$} to [out=-45,in=0] (t.east) ;
	\draw (n0b.-105) to [out=-135, in=90] ++ (-0.03,-0.03) node[zeroterm]{0};
	\draw (n0b.-75) to [out=-45, in=90] ++ (0.03,-0.03) node[zeroterm]{0};

	\node[right=0.2cm of dd1, scale=0.7] (g2) {
			\begin{quantikz}[column sep=5pt, row sep={0.45cm,between origins}, ampersand replacement=\&]
			\lstick{$q_2$} \&  \ctrl{2} \& \qw \\
			\lstick{$q_1$} \& \qw \& \qw\\
			\lstick{$q_0$} \&\gate{T} \& \qw
			\end{quantikz}
			};

	\node[right=-2mm of g2] (e2) {$\equiv$};
	
	\matrix[right=-1mm of e2, ampersand replacement=\&,every node/.style=vertex,row sep={0.4cm,between origins},column sep={0.2cm,between origins}] (dd) {
		\&\node (n2) {\small$q_2$}; \& \\
		\node (n1a) {\small$q_1$}; \& \& \node (n1b) {\small$q_1$}; \\
		\node (n0a) {\small$q_0$}; \& \& \node (n0b) {\small$q_0$}; \\
		\& \node[terminal] (t){\footnotesize 1}; \& \\
	};
	\draw (n2.north) to ++ (0,0.15);
	\draw (n2.south west) to [out=-135,in=90] (n1a.north);
	\draw (n2.south east) to [out=-45,in=90] (n1b.north);
	\draw (n2.-105) to [out=-135, in=90] ++ (-0.03,-0.03) node[zeroterm]{0};
	\draw (n2.-75) to [out=-45, in=90] ++ (0.03,-0.03) node[zeroterm]{0};
	
	\draw (n1a.south west) to [out=-135,in=135] (n0a.north west);
	\draw (n1a.south east) to [out=-45,in=45] (n0a.north east);
	\draw (n1a.-105) to [out=-135, in=90] ++ (-0.03,-0.03) node[zeroterm]{0};
	\draw (n1a.-75) to [out=-45, in=90] ++ (0.03,-0.03) node[zeroterm]{0};
	
	\draw (n1b.south west) to [out=-135,in=135] (n0b.north west);
	\draw (n1b.south east) to [out=-45,in=45] (n0b.north east) ;
	\draw (n1b.-105) to [out=-135, in=90] ++ (-0.03,-0.03) node[zeroterm]{0};
	\draw (n1b.-75) to [out=-45, in=90] ++ (0.03,-0.03) node[zeroterm]{0};
	
	\draw (n0a.south west) to [out=-135,in=180] (t.west);
	\draw (n0a.south east) to [out=-45,in=90] (t.north);
	\draw (n0a.-105) to [out=-135, in=90] ++ (-0.03,-0.03) node[zeroterm]{0};
	\draw (n0a.-75) to [out=-45, in=90] ++ (0.03,-0.03) node[zeroterm]{0};
	
	\draw (n0b.south west) to [out=-135,in=90] (t.north);
	\draw (n0b.south east) node[below right,font=\small, scale=0.5] {$\omega$} to [out=-45,in=0] (t.east) ;
	\draw (n0b.-105) to [out=-135, in=90] ++ (-0.03,-0.03) node[zeroterm]{0};
	\draw (n0b.-75) to [out=-45, in=90] ++ (0.03,-0.03) node[zeroterm]{0};
	\end{tikzpicture}}
	\caption{Decision Diagrams for operations shown in Fig.~\ref{fig:h_matrix}---\ref{fig:t_matrix}}
	\label{fig:gate_dd}
\end{figure}
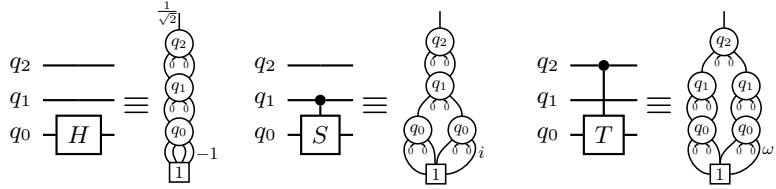

\begin{example}\label{ex:common_dd}
Consider again the quantum operations shown in Fig.~\ref{fig:operations}. Their functionalities can be represented efficiently in terms of Decision Diagrams as shown in~Fig.~\ref{fig:gate_dd}.
As can be seen, they allow for a rather compact representation ($3$--$5$ nodes vs. $8$--$16$ non-zero matrix entries).
\end{example}

Unfortunately, \emph{constructing} those representations for certain building blocks or even entire quantum algorithms can often not be conducted efficiently---even if it is conceptionally simple\footnote{The authors want to point out that this construction task is conceptionally different from and should not be confused with the classical simulation of quantum circuits which aims to calculate the resulting state vector for one particular input and not the complete functionality.}.
In fact, as reviewed in Section~\ref{sec:background}, the functionality of a quantum algorithm (given by a quantum circuit \mbox{$G=g_0,\dots,g_{m-1}$}) is described by the matrix \mbox{$U = U_{m-1} \cdots U_0$}, with $U_i$ being the matrix corresponding to gate~$g_i$ (for $0\leq i < m$). 
Hence, since the individual matrices~$U_i$ can usually be represented rather efficiently with either of the approaches reviewed above (arrays, tensor networks, DDs),  simply conducting multiplications on those representations should allow for an efficient construction of the entire functional representation.
But the more quantum operations are multiplied together, the more complex representations result---reducing the sparsity, increasing the degree of entanglement, or eliminating existing redundancies---and, hence, significantly slowing down the construction. Thus, while the multiplication operation itself is realized rather efficiently in general (utilizing, e.g., specialized techniques for sparse chain multiplication), the bottleneck arises from the consequences of consecutive multiplication.

\begin{figure}[t]
\centering
\resizebox{0.97\linewidth}{!}{
\begin{tikzpicture}
		
\node[scale=0.55, font=\tiny] (g0) {
			\begin{quantikz}[column sep=5pt, row sep={0.45cm,between origins}, ampersand replacement=\&]
			\&  \qw \& \qw \\
			\& \qw \& \qw\\
			\&\gate{H} \& \qw
			\end{quantikz}
			};
		
	\node[right = 0.1mm of g0, font=\tiny, scale=0.55] (g1) {
			\begin{quantikz}[column sep=5pt, row sep={0.45cm,between origins}, ampersand replacement=\&]
			\&  \qw \& \qw \\
			\& \ctrl{1} \& \qw\\
			\&\gate{S} \& \qw
			\end{quantikz}
			};
	
	\matrix[right = 0.1mm of g1,ampersand replacement=\&,every node/.style=vertex,row sep={0.4cm,between origins},column sep={0.15cm,between origins}] (dd1) {
		\&\node (n2) {\small$q_2$}; \& \\
		\&\node (n1) {\small$q_1$}; \& \\
		\node (n0a) {\small$q_0$}; \& \& \node (n0b) {\small$q_0$}; \\
		\& \node[terminal] (t){\footnotesize 1}; \& \\
	};
	\draw (n2.north) to ++ (0,0.15) node[left,inner sep=0pt,font=\small, scale=0.5] {$\tfrac{1}{\sqrt{2}}$};
	\draw (n2.south west) to [out=-135,in=135] (n1.north west);
	\draw (n2.south east) to [out=-45,in=45] (n1.north east);
	\draw (n2.-105) to [out=-135, in=90] ++ (-0.03,-0.03) node[zeroterm]{0};
	\draw (n2.-75) to [out=-45, in=90] ++ (0.03,-0.03) node[zeroterm]{0};
	
	\draw (n1.south west) to [out=-135,in=90] (n0a.north);
	\draw (n1.south east) to [out=-45,in=90] (n0b.north);
	\draw (n1.-105) to [out=-135, in=90] ++ (-0.03,-0.03) node[zeroterm]{0};
	\draw (n1.-75) to [out=-45, in=90] ++ (0.03,-0.03) node[zeroterm]{0};
	
	\draw (n0a.south west) to [out=-135,in=180] (t.west);
	\draw[densely dotted] (n0a.south east) to [out=-45,in=100] (t.100);
	\draw (n0a.-105) to [out=-105,in=155] (t.155);
	\draw (n0a.-75) to [out=-75,in=125] (t.125);
	
	\draw (n0b.south west) to [out=-135,in=80] (t.80);
	\draw[densely dotted, purple] (n0b.south east) to [out=-45,in=0] (t.east) ;
	\draw (n0b.-105) to [out=-105,in=55] (t.55);
	\draw[purple] (n0b.-75) to [out=-75,in=25] (t.25);
	
	\node[right = 0.1mm of dd1, font=\tiny, scale=0.55] (g2) {
			\begin{quantikz}[column sep=5pt, row sep={0.45cm,between origins}, ampersand replacement=\&]
			\&  \ctrl{2} \& \qw \\
			 \& \qw \& \qw\\
			\&\gate{T} \& \qw
			\end{quantikz}
			};	
	\matrix[right = 0.1mm of g2,ampersand replacement=\&,every node/.style=vertex,row sep={0.4cm,between origins},column sep={0.15cm,between origins}] (dd2) {
		\&\&\&\node (n2) {\small$q_2$}; \&\& \\
		\&\&\node (n1a) {\small$q_1$}; \& \& \node (n1b) {\small$q_1$};\& \\
		\node (n0a) {\small$q_0$}; \&\& \node (n0b) {\small$q_0$}; \& \& \node (n0c) {\small$q_0$}; \&\& \node (n0d) {\small$q_0$};\\
		\&\&\& \node[terminal] (t){\footnotesize 1}; \&\& \\
	};
	\draw (n2.north) to ++ (0,0.15) node[left,inner sep=0pt,font=\small, scale=0.5] {$\tfrac{1}{\sqrt{2}}$};
	\draw (n2.south west) to [out=-135,in=90] (n1a.north);
	\draw (n2.south east) to [out=-45,in=90] (n1b.north);
	\draw (n2.-105) to [out=-135, in=90] ++ (-0.03,-0.03) node[zeroterm]{0};
	\draw (n2.-75) to [out=-45, in=90] ++ (0.03,-0.03) node[zeroterm]{0};
	
	\draw (n1a.south west) to [out=-135,in=90] (n0a.north);
	\draw (n1a.-105) to [out=-135, in=90] ++ (-0.03,-0.03) node[zeroterm]{0};
	\draw (n1a.-75) to [out=-45, in=90] ++ (0.03,-0.03) node[zeroterm]{0};
	\draw (n1a.south east) to [out=-45,in=45] (n0b.45);
	
	\draw (n1b.south west) to [out=-135,in=135] (n0c.135);
	\draw (n1b.-105) to [out=-135, in=90] ++ (-0.03,-0.03) node[zeroterm]{0};
	\draw (n1b.-75) to [out=-45, in=90] ++ (0.03,-0.03) node[zeroterm]{0};
	\draw (n1b.south east) to [out=-45,in=90] (n0d.north);
	
	\draw (n0a.south west) to [out=-135,in=200] (t.200);
	\draw (n0a.-105) to [out=-105,in=190] (t.190);
	\draw (n0a.-75) to [out=-75,in=180] (t.180);
	\draw[densely dotted] (n0a.south east) to [out=-45,in=170] (t.170);
	
	\draw (n0b.south west) to [out=-135,in=155] (t.155);
	\draw (n0b.-105) to [out=-105,in=145] (t.145);
	\draw[purple] (n0b.-75) to [out=-75,in=135] (t.135);
	\draw[densely dotted,purple] (n0b.south east) to [out=-45,in=125] (t.125);
	
	\draw[thick] (n0c.south west) to [out=-135,in=55] (t.55);
	\draw[thick] (n0c.-105) to [out=-105,in=45] (t.45);
	\draw[blue, thick] (n0c.-75) to [out=-75,in=35] (t.35);
	\draw[densely dotted, blue, thick] (n0c.south east) to [out=-45,in=25] (t.25);
	
	\draw[thick] (n0d.south west) to [out=-135,in=10] (t.10);
	\draw[thick] (n0d.-105) to [out=-105,in=0] (t.0);
	\draw[red, thick] (n0d.-75) to [out=-75,in=-10] (t.-10);
	\draw[densely dotted, red, thick] (n0d.south east) to [out=-45,in=-20] (t.-20);

	\node[right = 0.1mm of dd2, font=\tiny, scale=0.55] (g3) {
			\begin{quantikz}[column sep=5pt, row sep={0.45cm,between origins}, ampersand replacement=\&]
			\&  \qw \& \qw \\
			\& \gate{H} \& \qw\\
			\&\qw \& \qw
			\end{quantikz}
	};
	
	\matrix[right = 0.1mm of g3,ampersand replacement=\&,every node/.style=vertex,row sep={0.4cm,between origins},column sep={0.15cm,between origins}] (dd3) {
		\&\&\&\node (n2) {\small$q_2$}; \&\& \\
		\&\&\node (n1a) {\small$q_1$}; \& \& \node (n1b) {\small$q_1$};\& \\
		\node (n0a) {\small$q_0$}; \&\& \node (n0b) {\small$q_0$}; \& \& \node (n0c) {\small$q_0$}; \&\& \node (n0d) {\small$q_0$};\\
		\&\&\& \node[terminal] (t){\footnotesize 1}; \&\& \\
	};
	\draw[thick] (n2.north) to ++ (0,0.15) node[left,inner sep=0pt,font=\small, scale=0.5] {$\tfrac{1}{\sqrt{4}}$};
	\draw (n2.south west) to [out=-135,in=90] (n1a.north);
	\draw (n2.south east) to [out=-45,in=90] (n1b.north);
	\draw (n2.-105) to [out=-135, in=90] ++ (-0.03,-0.03) node[zeroterm]{0};
	\draw (n2.-75) to [out=-45, in=90] ++ (0.03,-0.03) node[zeroterm]{0};
	
	\draw (n1a.south west) to [out=-135,in=90] (n0a.north);
	\draw[thick] (n1a.-105) to [out=-105, in=105] (n0b.105);
	\draw[thick] (n1a.-75) to [out=-75,in=75] (n0a.75);
	\draw[densely dotted, thick] (n1a.south east) to [out=-45,in=90] (n0b.north);
	
	\draw (n1b.south west) to [out=-135,in=90] (n0c.north);
	\draw[thick] (n1b.-105) to [out=-105, in=105] (n0d.105);
	\draw[thick] (n1b.-75) to [out=-75,in=75] (n0c.75);
	\draw[densely dotted, thick] (n1b.south east) to [out=-45,in=90] (n0d.north);
	
	\draw (n0a.south west) to [out=-135,in=200] (t.200);
	\draw (n0a.-105) to [out=-105,in=190] (t.190);
	\draw (n0a.-75) to [out=-75,in=180] (t.180);
	\draw[densely dotted] (n0a.south east) to [out=-45,in=170] (t.170);
	
	\draw (n0b.south west) to [out=-135,in=155] (t.155);
	\draw (n0b.-105) to [out=-105,in=145] (t.145);
	\draw[purple] (n0b.-75) to [out=-75,in=135] (t.135);
	\draw[densely dotted,purple] (n0b.south east) to [out=-45,in=125] (t.125);
	
	\draw (n0c.south west) to [out=-135,in=55] (t.55);
	\draw (n0c.-105) to [out=-105,in=45] (t.45);
	\draw[blue] (n0c.-75) to [out=-75,in=35] (t.35);
	\draw[densely dotted, blue] (n0c.south east) to [out=-45,in=25] (t.25);
	
	\draw (n0d.south west) to [out=-135,in=10] (t.10);
	\draw (n0d.-105) to [out=-105,in=0] (t.0);
	\draw[red] (n0d.-75) to [out=-75,in=-10] (t.-10);
	\draw[densely dotted, red] (n0d.south east) to [out=-45,in=-20] (t.-20);
	
	\node[right = 0.1mm of dd3, font=\tiny, scale=0.55] (g4) {
			\begin{quantikz}[column sep=5pt, row sep={0.45cm,between origins}, ampersand replacement=\&]
			\& \ctrl{1} \& \qw\\
			\&\gate{S} \& \qw\\
			\&  \qw \& \qw 
			\end{quantikz}
			};	
	\matrix[right = 0.1mm of g4, ampersand replacement=\&,every node/.style=vertex,row sep={0.4cm,between origins},column sep={0.15cm,between origins}] (dd4) {
		\&\&\&\node (n2) {\small$q_2$}; \&\& \\
		\&\&\node (n1a) {\small$q_1$}; \& \& \node (n1b) {\small$q_1$};\& \\
		\node (n0a) {\small$q_0$}; \&\& \node (n0b) {\small$q_0$}; \& \& \node (n0c) {\small$q_0$}; \&\& \node (n0d) {\small$q_0$};\\
		\&\&\& \node[terminal] (t){\footnotesize 1}; \&\& \\
	};
	\draw (n2.north) to ++ (0,0.15) node[left,inner sep=0pt,font=\small, scale=0.5] {$\tfrac{1}{\sqrt{4}}$};
	\draw (n2.south west) to [out=-135,in=90] (n1a.north);
	\draw (n2.south east) to [out=-45,in=90] (n1b.north);
	\draw (n2.-105) to [out=-135, in=90] ++ (-0.03,-0.03) node[zeroterm]{0};
	\draw (n2.-75) to [out=-45, in=90] ++ (0.03,-0.03) node[zeroterm]{0};
	
	\draw (n1a.south west) to [out=-135,in=90] (n0a.north);
	\draw (n1a.-105) to [out=-105, in=105] (n0b.105);
	\draw (n1a.-75) to [out=-75,in=75] (n0a.75);
	\draw[densely dotted] (n1a.south east) to [out=-45,in=90] (n0b.north);
	
	\draw (n1b.south west) to [out=-135,in=90] (n0c.north);
	\draw (n1b.-105) to [out=-105, in=105] (n0d.105);
	\draw[purple, thick] (n1b.-75) to [out=-75,in=75] (n0c.75);
	\draw[densely dotted, purple, thick] (n1b.south east) to [out=-45,in=90] (n0d.north);
	
	\draw (n0a.south west) to [out=-135,in=200] (t.200);
	\draw (n0a.-105) to [out=-105,in=190] (t.190);
	\draw (n0a.-75) to [out=-75,in=180] (t.180);
	\draw[densely dotted] (n0a.south east) to [out=-45,in=170] (t.170);
	
	\draw (n0b.south west) to [out=-135,in=155] (t.155);
	\draw (n0b.-105) to [out=-105,in=145] (t.145);
	\draw[purple] (n0b.-75) to [out=-75,in=135] (t.135);
	\draw[densely dotted,purple] (n0b.south east) to [out=-45,in=125] (t.125);
	
	\draw (n0c.south west) to [out=-135,in=55] (t.55);
	\draw (n0c.-105) to [out=-105,in=45] (t.45);
	\draw[blue] (n0c.-75) to [out=-75,in=35] (t.35);
	\draw[densely dotted, blue] (n0c.south east) to [out=-45,in=25] (t.25);
	
	\draw (n0d.south west) to [out=-135,in=10] (t.10);
	\draw (n0d.-105) to [out=-105,in=0] (t.0);
	\draw[red] (n0d.-75) to [out=-75,in=-10] (t.-10);
	\draw[densely dotted, red] (n0d.south east) to [out=-45,in=-20] (t.-20);
	
	\node[right = 0.1mm of dd4, font=\tiny, scale=0.55] (g5) {
			\begin{quantikz}[column sep=5pt, row sep={0.45cm,between origins}, ampersand replacement=\&]
			\&\gate{H} \& \qw\\
			\&  \qw \& \qw \\
			\& \qw \& \qw
			\end{quantikz}
			};	
	\matrix[right = 0.1mm of g5, ampersand replacement=\&,every node/.style=vertex,row sep={0.4cm,between origins},column sep={0.15cm,between origins}] (dd5) {
		\&\&\&\node (n2) {\small$q_2$}; \&\& \\
		\&\&\node (n1a) {\small$q_1$}; \& \& \node (n1b) {\small$q_1$};\& \\
		\node (n0a) {\small$q_0$}; \&\& \node (n0b) {\small$q_0$}; \& \& \node (n0c) {\small$q_0$}; \&\& \node (n0d) {\small$q_0$};\\
		\&\&\& \node[terminal] (t){\footnotesize 1}; \&\& \\
	};
	\draw[thick] (n2.north) to ++ (0,0.15) node[left,inner sep=0pt,font=\small, scale=0.5] {$\tfrac{1}{\sqrt{8}}$};
	\draw (n2.south west) to [out=-135,in=90] (n1a.north);
	\draw[thick] (n2.-105) to [out=-135, in=135] (n1b.135);
	\draw[thick] (n2.-75) to [out=-45, in=45] (n1a.45);
	\draw[densely dotted, thick] (n2.south east) to [out=-45,in=90] (n1b.north);

	\draw (n1a.south west) to [out=-135,in=90] (n0a.north);
	\draw (n1a.-105) to [out=-105, in=105] (n0b.105);
	\draw (n1a.-75) to [out=-75,in=75] (n0a.75);
	\draw[densely dotted] (n1a.south east) to [out=-45,in=90] (n0b.north);
	
	\draw (n1b.south west) to [out=-135,in=90] (n0c.north);
	\draw (n1b.-105) to [out=-105, in=105] (n0d.105);
	\draw[purple] (n1b.-75) to [out=-75,in=75] (n0c.75);
	\draw[densely dotted, purple] (n1b.south east) to [out=-45,in=90] (n0d.north);
	
	\draw (n0a.south west) to [out=-135,in=200] (t.200);
	\draw (n0a.-105) to [out=-105,in=190] (t.190);
	\draw (n0a.-75) to [out=-75,in=180] (t.180);
	\draw[densely dotted] (n0a.south east) to [out=-45,in=170] (t.170);
	
	\draw (n0b.south west) to [out=-135,in=155] (t.155);
	\draw (n0b.-105) to [out=-105,in=145] (t.145);
	\draw[purple] (n0b.-75) to [out=-75,in=135] (t.135);
	\draw[densely dotted,purple] (n0b.south east) to [out=-45,in=125] (t.125);
	
	\draw (n0c.south west) to [out=-135,in=55] (t.55);
	\draw (n0c.-105) to [out=-105,in=45] (t.45);
	\draw[blue] (n0c.-75) to [out=-75,in=35] (t.35);
	\draw[densely dotted, blue] (n0c.south east) to [out=-45,in=25] (t.25);
	
	\draw (n0d.south west) to [out=-135,in=10] (t.10);
	\draw (n0d.-105) to [out=-105,in=0] (t.0);
	\draw[red] (n0d.-75) to [out=-75,in=-10] (t.-10);
	\draw[densely dotted, red] (n0d.south east) to [out=-45,in=-20] (t.-20);

	\draw[-Latex, thick, font=\tiny] ($(g1) - (0.4,1)$) node[left, fill=white] {\textbf{Size:}}-- ($(g1) - (0,1)$) -- ($(dd1) - (0,1)$) node[fill=white]{$\mathbf{4}$} -- ($(dd2) - (0,1)$) node[fill=white]{$\mathbf{7}$} -- ($(dd3) - (0,1)$) node[fill=white]{$\mathbf{7}$} -- ($(dd4) - (0,1)$) node[fill=white]{$\mathbf{7}$}-- ($(dd5) - (0,1)$) node[fill=white]{$\mathbf{7}$} -- ($(dd5) - (-0.7,1)$);
	
	\node (m1) at ($(g0)!0.5!(g1) +(0.04,0.06)$) {$\mathbf{\times}$};
	
	\draw[thick] ($(g1.north)!0.35!(dd1.north)$) to [out=-45, in=135] (dd1.west);
	\draw[thick] ($(g1.south)!0.35!(dd1.south)$) to [out=45, in=-135] (dd1.west);
	
	\node (m2) at ($(dd1)!0.56!(g2) + (0.0,0.04)$) {$\mathbf{\times}$};
	
	\draw[thick] ($(g2.north)!0.25!(dd2.north)$) to [out=-45, in=135] (dd2.west);
	\draw[thick] ($(g2.south)!0.25!(dd2.south)$) to [out=45, in=-135] (dd2.west);
	
	\node (m3) at ($(dd2)!0.62!(g3)$) {$\mathbf{\times}$};
	
	\draw[thick] ($(g3.north)!0.25!(dd3.north)$) to [out=-45, in=135] (dd3.west);
	\draw[thick] ($(g3.south)!0.25!(dd3.south)$) to [out=45, in=-135] (dd3.west);
	
	\node (m4) at ($(dd3)!0.62!(g4)+ (0.02, -0.01)$) {$\mathbf{\times}$};
	
	\draw[thick] ($(g4.north)!0.25!(dd4.north)$) to [out=-45, in=135] (dd4.west);
	\draw[thick] ($(g4.south)!0.25!(dd4.south)$) to [out=45, in=-135] (dd4.west);
	
	\node (m5) at ($(dd4)!0.64!(g5) + (0.02, -0.04)$) {$\mathbf{\times}$};
	
	\draw[thick] ($(g5.north)!0.25!(dd5.north)$) to [out=-45, in=135] (dd5.west);
	\draw[thick] ($(g5.south)!0.25!(dd5.south)$) to [out=45, in=-135] (dd5.west);
	
\end{tikzpicture}}\vspace*{-3mm}
\caption{State-of-the-art DD composition sequence for 3-qubit QFT (see~Fig.~\ref{fig:qft3})}
\label{fig:reference_sequence}
\vspace*{-2em}
\end{figure}
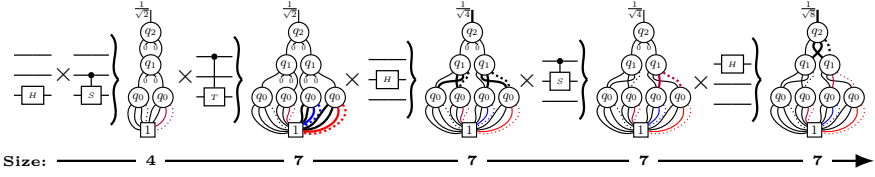

\begin{example}\label{ex:qft_sota}
Consider again the circuit for the 3-qubit QFT from Fig.~\ref{fig:qft3}.
Constructing its functionality requires multiplying the individual representations of all $6$ gates.
The multiplication of two Decision Diagrams (representing matrices $U$ and $V$) is recursively broken down into sub-expressions according to
\begin{equation*}
\begin{bNiceMatrix}[]
U_{00} & U_{01} \\ U_{10} & U_{11}
\end{bNiceMatrix}\cdot
\begin{bNiceMatrix}[]
V_{00} & V_{01} \\ V_{10} & V_{11}
\end{bNiceMatrix} =
\begin{bNiceMatrix}[]
(U_{00}V_{00} + U_{01}V_{10}) & (U_{00}V_{01} + U_{01}V_{11}) \\ (U_{10}V_{00} + U_{11}V_{10}) & (U_{10}V_{01} + U_{11}V_{11})
\end{bNiceMatrix},
\end{equation*}
until only operations on complex numbers remain. That results in a complexity which scales with the product of the number of nodes in the Decision Diagrams to be multiplied.
Carrying out all multiplications results in the evolution of representations as shown in~Fig.~\ref{fig:reference_sequence}\footnote{Different edge weights are indicated by dotted (\mbox{$\equiv$ negative}) and/or colored ($\equiv$ 1, {\color{purple}$i$}, {\color{red}$\omega$} and~{\color{blue}$\omega^3$}) lines. This suffices to illustrate the evolution of the Decision Diagrams' size, i.e., their node count.}.
While, as already shown by means of Fig.~\ref{fig:gate_dd}, the functionality of single operations can be represented compactly, the multiplication needed to construct the overall functionality quickly increases the complexity. In fact, after two multiplications, further computations have to be conducted on a representation as large as the final result. 
\end{example}

Evaluations on larger examples than the one above confirm that, in many cases, we may be able to represent (i.e.,~store) the overall functionality of certain building blocks or an entire quantum algorithm (and use it for tasks such as synthesis/compilation, simulation, or verification), but we may not be able to construct this representation in feasible time. This constitutes a severe bottleneck for many applications in the domain of quantum computing.

\section{Proposed Approaches}
\label{sec:proposed}
In order to overcome the bottleneck discussed in the previous section, we propose to approach the construction of functional representations for building blocks or entire quantum algorithms with different strategies. 
We distinguish thereby two use cases: First, a general construction scheme is presented which can be applied for arbitrary functionality. Afterwards, we present a second scheme which is dedicated to repeating structures as they frequently occur in quantum algorithms (e.g., by means of Grover iterations or quantum walks). The resulting schemes allow to speed up the construction of the desired
functional representation considerably and even manage to complete the construction where existing methods time out.

\subsection{General Scheme}
\label{sec:general}

The observations from Section~\ref{sec:building} show that the bottleneck emerges as a result of a large number of matrix-matrix multiplications on rather large representations. 
Hence, in order to avoid this, we propose to conduct as many of those multiplications on as small as possible representations, e.g.,~on the original gate representations. 
Here, the fact that \mbox{matrix} multiplication is associative comes in handy as it allows to conduct those multiplications in a different order.

More precisely, assume, for sake of simplicity, that the number~$m$ of operations of a given building block or quantum algorithm is a power of two, i.e., $m=2^k$ (for some $k\in\mathbb{N}$).
Then, grouping the set of $m$ operations into $m/2$ consecutive pairs, i.e., 
\[
(U_{m-1}\cdot U_{m-2})\cdot\hdots\cdot(U_3\cdot U_2)\cdot(U_1\cdot U_0) = U,
\]
and performing the pairwise multiplications $(U_{i+1}\cdot U_i) = U_{i+1,i}$, leaves $m/2 = 2^{k-1}$ factors to be multiplied, i.e., 
\[
U_{m-1,m-2} \cdot\hdots\cdot U_{1,0} = U.
\]
Recursively applying this idea eventually results in the construction of the full functional representation \mbox{$U\equiv U_{m-1,\dots,0}$}---requiring a total of $k$ levels of pairwise grouping and multiplication.
In case $m$ is not a power of two, in some levels a pair may ``degenerate'' to a single operation.

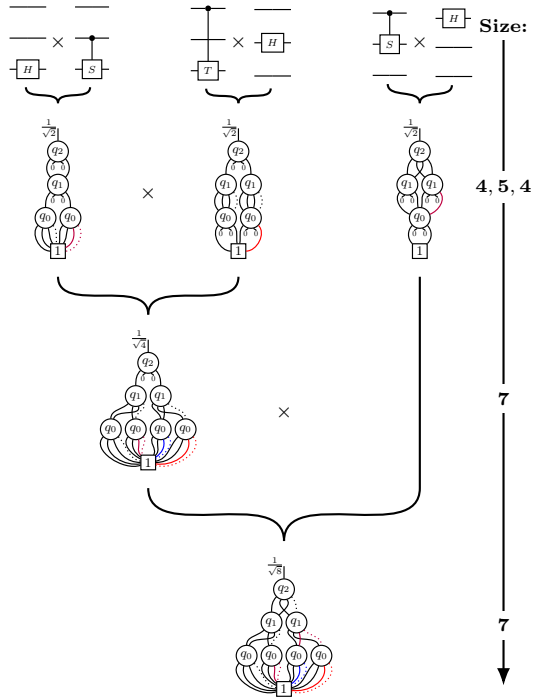
\begin{figure}[t]
\centering
\resizebox{0.6\linewidth}{!}{
\begin{tikzpicture}
\node[font=\scriptsize, scale=0.55] (g0) {
			\begin{quantikz}[column sep=5pt, row sep={0.75cm,between origins}, ampersand replacement=\&]
			\&  \qw \& \qw \\
			\& \qw \& \qw\\
			\&\gate{H} \& \qw
			\end{quantikz}
			};
			
\node[right = 1mm of g0, font=\scriptsize, scale=0.55] (g1) {
			\begin{quantikz}[column sep=5pt, row sep={0.75cm,between origins}, ampersand replacement=\&]
			\&  \qw \& \qw \\
			 \& \ctrl{1} \& \qw\\
			 \&\gate{S} \& \qw
			\end{quantikz}
			};

\node[right = 0.8cm of g1, font=\scriptsize, scale=0.55] (g2) {
			\begin{quantikz}[column sep=5pt, row sep={0.75cm,between origins}, ampersand replacement=\&]
			\&  \ctrl{2} \& \qw \\
			\& \qw \& \qw\\
			\&\gate{T} \& \qw
			\end{quantikz}
			};

\node[right = 1mm of g2, font=\scriptsize, scale=0.55] (g3) {
			\begin{quantikz}[column sep=5pt, row sep={0.8cm,between origins}, ampersand replacement=\&]
			\&  \qw \& \qw \\
			\& \gate{H} \& \qw\\
			\&\qw \& \qw
			\end{quantikz}
	};

\node[right = 0.8cm of g3, font=\scriptsize, scale=0.55] (g4) {
			\begin{quantikz}[column sep=5pt, row sep={0.75cm,between origins}, ampersand replacement=\&]
			\& \ctrl{1} \& \qw\\
			\&\gate{S} \& \qw\\
			\&  \qw \& \qw 
			\end{quantikz}
			};	

\node[right = 1mm of g4, font=\scriptsize, scale=0.55] (g5) {
			\begin{quantikz}[column sep=5pt, row sep={0.7cm,between origins}, ampersand replacement=\&]
			\&\gate{H} \& \qw\\
			\&  \qw \& \qw \\
			\& \qw \& \qw
			\end{quantikz}
			};	

\node (dd1) at ($(g0)!0.5!(g1)-(0,2.0)$) {
\scalebox{1.1}{\begin{tikzpicture}
\matrix[ampersand replacement=\&,every node/.style=vertex,row sep={0.4cm,between origins},column sep={0.15cm,between origins}]  {
	\&\node (n2) {\small$q_2$}; \& \\
	\&\node (n1) {\small$q_1$}; \& \\
	\node (n0a) {\small$q_0$}; \& \& \node (n0b) {\small$q_0$}; \\
	\& \node[terminal] (t){\footnotesize 1}; \& \\
};
\draw (n2.north) to ++ (0,0.15) node[left,inner sep=0pt,font=\small, scale=0.5] {$\tfrac{1}{\sqrt{2}}$};
\draw (n2.south west) to [out=-135,in=135] (n1.north west);
\draw (n2.south east) to [out=-45,in=45] (n1.north east);
\draw (n2.-105) to [out=-135, in=90] ++ (-0.03,-0.03) node[zeroterm]{0};
\draw (n2.-75) to [out=-45, in=90] ++ (0.03,-0.03) node[zeroterm]{0};

\draw (n1.south west) to [out=-135,in=90] (n0a.north);
\draw (n1.south east) to [out=-45,in=90] (n0b.north);
\draw (n1.-105) to [out=-135, in=90] ++ (-0.03,-0.03) node[zeroterm]{0};
\draw (n1.-75) to [out=-45, in=90] ++ (0.03,-0.03) node[zeroterm]{0};

\draw (n0a.south west) to [out=-135,in=180] (t.west);
\draw[densely dotted] (n0a.south east) to [out=-45,in=100] (t.100);
\draw (n0a.-105) to [out=-105,in=155] (t.155);
\draw (n0a.-75) to [out=-75,in=125] (t.125);

\draw (n0b.south west) to [out=-135,in=80] (t.80);
\draw[densely dotted, purple] (n0b.south east) to [out=-45,in=0] (t.east) ;
\draw (n0b.-105) to [out=-105,in=55] (t.55);
\draw[purple] (n0b.-75) to [out=-75,in=25] (t.25);
\end{tikzpicture}}};

\node (dd2) at ($(g2)!0.5!(g3)-(0,2.0)$) {
\scalebox{1.1}{\begin{tikzpicture}
\matrix[ampersand replacement=\&,every node/.style=vertex,row sep={0.4cm,between origins},column sep={0.15cm,between origins}] {
	\&\node (n2) {\small$q_2$}; \& \\
	\node (n1a) {\small$q_1$}; \& \& \node (n1b) {\small$q_1$}; \\
	\node (n0a) {\small$q_0$}; \& \& \node (n0b) {\small$q_0$}; \\
	\& \node[terminal] (t){\footnotesize 1}; \& \\
};
\draw (n2.north) to ++ (0,0.15) node[left,inner sep=0pt,font=\small, scale=0.5] {$\tfrac{1}{\sqrt{2}}$};
\draw (n2.south west) to [out=-135,in=90] (n1a.north);
\draw (n2.-105) to [out=-105, in=90] ++ (-0.03,-0.03) node[zeroterm]{0};
\draw (n2.-75) to [out=-75, in=90] ++ (0.03,-0.03) node[zeroterm]{0};
\draw (n2.south east) to [out=-45,in=90] (n1b.90);

\draw (n1a.-135) to [out=-135,in=135] (n0a.135);
\draw (n1a.-105) to [out=-105,in=105] (n0a.105);
\draw (n1a.-75) to [out=-75,in=75] (n0a.75);
\draw[densely dotted] (n1a.-45) to [out=-45,in=45] (n0a.45);

\draw (n1b.-135) to [out=-135,in=135] (n0b.135);
\draw (n1b.-105) to [out=-105,in=105] (n0b.105);
\draw (n1b.-75) to [out=-75,in=75] (n0b.75);
\draw[densely dotted] (n1b.-45) to [out=-45,in=45] (n0b.45);

\draw (n0a.south west) to [out=-135,in=180] (t.west);
\draw (n0a.-105) to [out=-135, in=90] ++ (-0.03,-0.03) node[zeroterm]{0};
\draw (n0a.-75) to [out=-75, in=90] ++ (0.03,-0.03) node[zeroterm]{0};
\draw (n0a.south east) to [out=-45,in=100] (t.100);

\draw (n0b.south west) to [out=-135, in=80] (t.80);
\draw (n0b.-105) to [out=-105, in=90] ++ (-0.03,-0.03) node[zeroterm]{0};
\draw (n0b.-75) to [out=-75, in=90] ++ (0.03,-0.03) node[zeroterm]{0};
\draw[red] (n0b.south east) to [out=-45,in=0] (t.east) ;
\end{tikzpicture}}};

\node (dd3) at ($(g4)!0.5!(g5)-(0,2.0)$) {
\scalebox{1.1}{\begin{tikzpicture}
\matrix[ampersand replacement=\&,every node/.style=vertex,row sep={0.4cm,between origins},column sep={0.15cm,between origins}] {
	\&\node (n2) {\small$q_2$}; \& \\
	\node (n1a) {\small$q_1$}; \& \& \node (n1b) {\small$q_1$}; \\
	\& \node (n0) {\small$q_0$}; \\
	\& \node[terminal] (t){\footnotesize 1}; \& \\
};
\draw (n2.north) to ++ (0,0.15) node[left,inner sep=0pt,font=\small, scale=0.5] {$\tfrac{1}{\sqrt{2}}$};
\draw (n2.south west) to [out=-135,in=90] (n1a.north);
\draw (n2.-105) to [out=-105, in=135] (n1b.135);
\draw (n2.-75) to [out=-75, in=45] (n1a.45);
\draw (n2.south east) to [out=-45,in=90] (n1b.90);

\draw (n1a.-135) to [out=-135,in=160] (n0.160);
\draw (n1a.-105) to [out=-105, in=90] ++ (-0.03,-0.03) node[zeroterm]{0};
\draw (n1a.-75) to [out=-75, in=90] ++ (0.03,-0.03) node[zeroterm]{0};
\draw (n1a.-45) to [out=-45,in=100] (n0.100);

\draw (n1b.-135) to [out=-135,in=80] (n0.80);
\draw (n1b.-105) to [out=-105, in=90] ++ (-0.03,-0.03) node[zeroterm]{0};
\draw (n1b.-75) to [out=-75, in=90] ++ (0.03,-0.03) node[zeroterm]{0};
\draw[purple] (n1b.-45) to [out=-45,in=20] (n0.20);

\draw (n0.south west) to [out=-135,in=135] (t.north west);
\draw (n0.-105) to [out=-105, in=90] ++ (-0.03,-0.03) node[zeroterm]{0};
\draw (n0.-75) to [out=-75, in=90] ++ (0.03,-0.03) node[zeroterm]{0};
\draw (n0.south east) to [out=-45,in=45] (t.north east);
\end{tikzpicture}}};

\node (dd4) at ($(dd1)!0.5!(dd2)-(0,2.8)$) {
\scalebox{1.1}{\begin{tikzpicture}
\matrix[ampersand replacement=\&,every node/.style=vertex,row sep={0.4cm,between origins},column sep={0.15cm,between origins}]  {
	\&\&\&\node (n2) {\small$q_2$}; \&\& \\
	\&\&\node (n1a) {\small$q_1$}; \& \& \node (n1b) {\small$q_1$};\& \\
	\node (n0a) {\small$q_0$}; \&\& \node (n0b) {\small$q_0$}; \& \& \node (n0c) {\small$q_0$}; \&\& \node (n0d) {\small$q_0$};\\
	\&\&\& \node[terminal] (t){\footnotesize 1}; \&\& \\
};
\draw (n2.north) to ++ (0,0.15) node[left,inner sep=0pt,font=\small, scale=0.5] {$\tfrac{1}{\sqrt{4}}$};
\draw (n2.south west) to [out=-135,in=90] (n1a.north);
\draw (n2.south east) to [out=-45,in=90] (n1b.north);
\draw (n2.-105) to [out=-135, in=90] ++ (-0.03,-0.03) node[zeroterm]{0};
\draw (n2.-75) to [out=-45, in=90] ++ (0.03,-0.03) node[zeroterm]{0};

\draw (n1a.south west) to [out=-135,in=90] (n0a.north);
\draw (n1a.-105) to [out=-105, in=105] (n0b.105);
\draw (n1a.-75) to [out=-75,in=75] (n0a.75);
\draw[densely dotted] (n1a.south east) to [out=-45,in=90] (n0b.north);

\draw (n1b.south west) to [out=-135,in=90] (n0c.north);
\draw (n1b.-105) to [out=-105, in=105] (n0d.105);
\draw (n1b.-75) to [out=-75,in=75] (n0c.75);
\draw[densely dotted] (n1b.south east) to [out=-45,in=90] (n0d.north);

\draw (n0a.south west) to [out=-135,in=200] (t.200);
\draw (n0a.-105) to [out=-105,in=190] (t.190);
\draw (n0a.-75) to [out=-75,in=180] (t.180);
\draw[densely dotted] (n0a.south east) to [out=-45,in=170] (t.170);

\draw (n0b.south west) to [out=-135,in=155] (t.155);
\draw (n0b.-105) to [out=-105,in=145] (t.145);
\draw[purple] (n0b.-75) to [out=-75,in=135] (t.135);
\draw[densely dotted,purple] (n0b.south east) to [out=-45,in=125] (t.125);

\draw (n0c.south west) to [out=-135,in=55] (t.55);
\draw (n0c.-105) to [out=-105,in=45] (t.45);
\draw[blue] (n0c.-75) to [out=-75,in=35] (t.35);
\draw[densely dotted, blue] (n0c.south east) to [out=-45,in=25] (t.25);

\draw (n0d.south west) to [out=-135,in=10] (t.10);
\draw (n0d.-105) to [out=-105,in=0] (t.0);
\draw[red] (n0d.-75) to [out=-75,in=-10] (t.-10);
\draw[densely dotted, red] (n0d.south east) to [out=-45,in=-20] (t.-20);
\end{tikzpicture}}};

\node (dd5) at ($(dd4)!0.5!(dd3)-(0mm,4.4)$) {
\scalebox{1.1}{\begin{tikzpicture}
\matrix[ampersand replacement=\&,every node/.style=vertex,row sep={0.4cm,between origins},column sep={0.15cm,between origins}]  {
	\&\&\&\node (n2) {\small$q_2$}; \&\& \\
	\&\&\node (n1a) {\small$q_1$}; \& \& \node (n1b) {\small$q_1$};\& \\
	\node (n0a) {\small$q_0$}; \&\& \node (n0b) {\small$q_0$}; \& \& \node (n0c) {\small$q_0$}; \&\& \node (n0d) {\small$q_0$};\\
	\&\&\& \node[terminal] (t){\footnotesize 1}; \&\& \\
};
\draw (n2.north) to ++ (0,0.15) node[left,inner sep=0pt,font=\small, scale=0.5] {$\tfrac{1}{\sqrt{8}}$};
\draw (n2.south west) to [out=-135,in=90] (n1a.north);
\draw (n2.-105) to [out=-135, in=135] (n1b.135);
\draw (n2.-75) to [out=-45, in=45] (n1a.45);
\draw[densely dotted] (n2.south east) to [out=-45,in=90] (n1b.north);

\draw (n1a.south west) to [out=-135,in=90] (n0a.north);
\draw (n1a.-105) to [out=-105, in=105] (n0b.105);
\draw (n1a.-75) to [out=-75,in=75] (n0a.75);
\draw[densely dotted] (n1a.south east) to [out=-45,in=90] (n0b.north);

\draw (n1b.south west) to [out=-135,in=90] (n0c.north);
\draw (n1b.-105) to [out=-105, in=105] (n0d.105);
\draw[purple] (n1b.-75) to [out=-75,in=75] (n0c.75);
\draw[densely dotted, purple] (n1b.south east) to [out=-45,in=90] (n0d.north);

\draw (n0a.south west) to [out=-135,in=200] (t.200);
\draw (n0a.-105) to [out=-105,in=190] (t.190);
\draw (n0a.-75) to [out=-75,in=180] (t.180);
\draw[densely dotted] (n0a.south east) to [out=-45,in=170] (t.170);

\draw (n0b.south west) to [out=-135,in=155] (t.155);
\draw (n0b.-105) to [out=-105,in=145] (t.145);
\draw[purple] (n0b.-75) to [out=-75,in=135] (t.135);
\draw[densely dotted,purple] (n0b.south east) to [out=-45,in=125] (t.125);

\draw (n0c.south west) to [out=-135,in=55] (t.55);
\draw (n0c.-105) to [out=-105,in=45] (t.45);
\draw[blue] (n0c.-75) to [out=-75,in=35] (t.35);
\draw[densely dotted, blue] (n0c.south east) to [out=-45,in=25] (t.25);

\draw (n0d.south west) to [out=-135,in=10] (t.10);
\draw (n0d.-105) to [out=-105,in=0] (t.0);
\draw[red] (n0d.-75) to [out=-75,in=-10] (t.-10);
\draw[densely dotted, red] (n0d.south east) to [out=-45,in=-20] (t.-20);
\end{tikzpicture}}};

\draw[thick] (g0.south) to [out=-90, in=90] ($(dd1.north)+(0,1mm)$);
\draw[thick] (g1.south) to [out=-90, in=90] ($(dd1.north)+(0,1mm)$);
\draw[thick] (g2.south) to [out=-90, in=90] ($(dd2.north)+(0,1mm)$);
\draw[thick] ($(g3.south)-(0,0.5mm)$) to [out=-90, in=90] ($(dd2.north)+(0,1mm)$);
\draw[thick] (g4.south) to [out=-90, in=90] ($(dd3.north)+(0,1mm)$);
\draw[thick] (g5.south) to [out=-90, in=90] ($(dd3.north)+(0,1mm)$);

\draw[thick] ($(dd1.south)$) to [out=-90, in=90] ($(dd4.north)+(0,1mm)$);
\draw[thick] ($(dd2.south)$) to [out=-90, in=90] ($(dd4.north)+(0,1mm)$);

\draw[thick] ($(dd4.south)$) to [out=-90, in=90] ($(dd5.north)+(0,1mm)$);
\draw[thick] ($(dd3.south)$) to ($(dd3.south)-(0,2.8cm)$) to [out=-90, in=90] ($(dd5.north)+(0,1mm)$);

\node[font=\scriptsize] (m1) at ($(g0)!0.5!(g1)$) {$\mathbf{\times}$};
\node[font=\scriptsize] (m2) at ($(g2)!0.5!(g3)$) {$\mathbf{\times}$};
\node[font=\scriptsize] (m3) at ($(g4)!0.5!(g5)$) {$\mathbf{\times}$};

\node[font=\scriptsize] (m4) at ($(dd1)!0.5!(dd2)$) {$\mathbf{\times}$};

\node[font=\scriptsize] (m5) at ($(dd4)!0.5!(dd3)+(0,-1.5)$) {$\mathbf{\times}$};

\draw[-Latex, thick] ($(g5.45)+(0.3,-0.15)$) node[fill=white, font=\scriptsize] (size) {\textbf{Size:}} -- ($(size)+(0.0,-2.15)$) node[fill=white, font=\scriptsize] (fff) {$\mathbf{4,5,4}$}-- ($(fff)+(0,-2.8)$) node[fill=white, font=\scriptsize] (seven) {$\mathbf{7}$}-- ($(seven)+(0,-3.0)$) node[fill=white, font=\scriptsize] (s2) {$\mathbf{7}$}-- ($(s2)+(0,-0.8)$);
\end{tikzpicture}}
\vspace*{-2mm}
\caption{Proposed approach applied to 3-qubit QFT}
\label{fig:proposed_qft}
\end{figure}

\begin{example}\label{ex:proposed_qft}
Consider again the circuit for the 3-qubit QFT from Fig.~\ref{fig:qft3}. 
Conducting the operations according to the proposed scheme results in the evolution of representations as sketched in Fig.~\ref{fig:proposed_qft}. 
As can be seen, this leads to a much more efficient construction compared to the current \mbox{state-of-the-art} method illustrated before in Example~\ref{ex:qft_sota}:
While, thus far, the multiplications resulted in intermediate representations with $4$, $7$, $7$, $7$, and $7$~nodes (see~Fig.~\ref{fig:reference_sequence}), now the construction results in Decision Diagrams with $4$, $5$, and $4$~nodes (first level), $7$~nodes (second level), as well as $7$~nodes (third level).
While the total number of operations (as well as the final result) is obviously the same, 
more matrix-matrix multiplications are conducted on smaller representations.
Furthermore, while, for small examples as considered here, this difference might seem negligible, evaluations on larger quantum algorithms show that this change in the order of multiplications has a substantial effect on the efficiency of the construction.
\end{example}

In general, employing the proposed scheme creates a \mbox{tree-like} hierarchy of matrix compositions.
In each level $l\in\{0,\dots,k\}$, at most $2^l$ operations contribute to a specific group of compositions.
As a consequence, the intermediate functionalities during the construction can frequently be represented in a much more compact fashion (since these remain rather compact and/or sparse in many cases)---leading to fewer multiplications involving large representations.

Clearly, associativity of matrix multiplication allows for partitioning schemes beyond pairwise grouping.
Determining an optimal partitioning scheme can be related to finding an optimal contraction order of a tensor network (itself an \mbox{NP-hard} problem~\cite{chi-chungOptimizingClassMultidimensional1997}).
In this sense, the proposed scheme can be viewed as one possible heuristic of tackling the contraction problem for quantum circuits.

At a first glance, the proposed scheme merely trades runtime for space: many operations can be conducted on rather small intermediate representations. But, this requires to store a lot more intermediate results when compared to sequential approaches---specifically in the first level of the multiplication hierarchy, where $m/2$ Decision Diagrams have to be stored. 
However, as those \enquote{early} intermediate results correspond to circuits with very low depth, their representations are rather compact and frequently contain redundant subparts that can be shared\footnote{Decision Diagram packages, e.g., typically employ a unique table where all nodes are stored~\cite{zulehnerHowEfficientlyHandle2019}. Thus, even when multiple different Decision Diagrams are stored concurrently, sharing reduces the memory footprint considerably.}.
Moreover, the proposed approach can be realized using a stack for the intermediate results containing at most $\mathcal{O}(\log m)$ elements at any given time by proceeding in a depth-first fashion.

\subsection{Exploiting Repeating Structures}
\label{sec:special}

Besides the general scheme proposed above, the made observations and findings can further be tailored to repeating structures in quantum algorithms---allowing for even more improvements in the construction of functional representations. This is described in the following section.
To this end, recall that many quantum algorithms rely on repeated building blocks realizing a certain kind of iteration, e.g.,~Grover's search algorithm~\cite{groverFastQuantumMechanical1996}, Quantum Random Walks~\cite{douglasEfficientQuantumCircuit2009}, Amplitude Estimation~\cite{brassardQuantumAmplitudeAmplification2002}, or Phase Estimation~\cite{Kit96}.
Usually this type of algorithms consists of an initialization phase and an iteration phase comprised of multiple (identical) iteration steps.
 Inspired by emulation techniques~\cite{steigerProjectQOpenSource2018,zulehnerMatrixVectorVsMatrixMatrix2019}, the current state of the art accelerates the construction of the corresponding functional representation by
constructing a single initialization matrix~$U_{\mathit{init}}$ followed by \emph{multiple} multiplications with an iteration matrix~$U_{\mathit{iter}}$ (which has to be constructed only once), i.e.,
$(U_{\mathit{iter}}\cdot\hdots\cdot U_{\mathit{iter}}) U_{\mathit{init}} = U$.

This procedure can be drastically improved further by,
first, efficiently constructing the individual representations for $U_{\mathit{init}}$ and $U_{\mathit{iter}}$ using the general scheme proposed in Section~\ref{sec:general} and,
then, employing a binary exponentiation scheme for the sequence of multiplications involving $U_{\mathit{iter}}$. 
Assume, for sake of simplicity, that the number of iterations $N$ is a power of two, i.e., $N=2^k$ for some $k\in\mathbb{N}$.
 Then,\[
\overbrace{(U_{\mathit{iter}}\cdots U_{\mathit{iter}})}^{\mbox{$N$}} U_{\mathit{init}} = \overbrace{(U_{\mathit{iter}}^2\cdots U_{\mathit{iter}}^2)}^{\mbox{$N/2$}} U_{\mathit{init}} = \ldots = U_{iter}^N U_{init} = U.
\]
More precisely, once the iteration matrix $U_{\mathit{iter}}$ has been efficiently constructed, it is sufficient to carry out only one multiplication $U_{\mathit{iter}}^{2^l} \cdot U_{\mathit{iter}}^{2^l} = U_{\mathit{iter}}^{2^{l+1}}$ (squaring the current representation)
at each level $l\in\{0,\dots,k-1\}$, since all other multiplications are going to have the same result.
As a last step, the initialization matrix $U_{\mathit{init}}$ is multiplied to $U_{iter}^N$---yielding the desired representation~$U$.
Hence, only $k=\log_2(N)+1$ (instead of $N$) building block multiplications are required to construct the 
representation---an exponential reduction.
In case~$N$ is not a power of two, at most one additional multiplication per level is necessary. Thus, even in this case $\mathcal{O}(\log N)$ building block multiplications are sufficient for the construction.
As a matter of fact, this does not just reduce the number of multiplications exponentially, it also avoids many computations on potentially large representations.

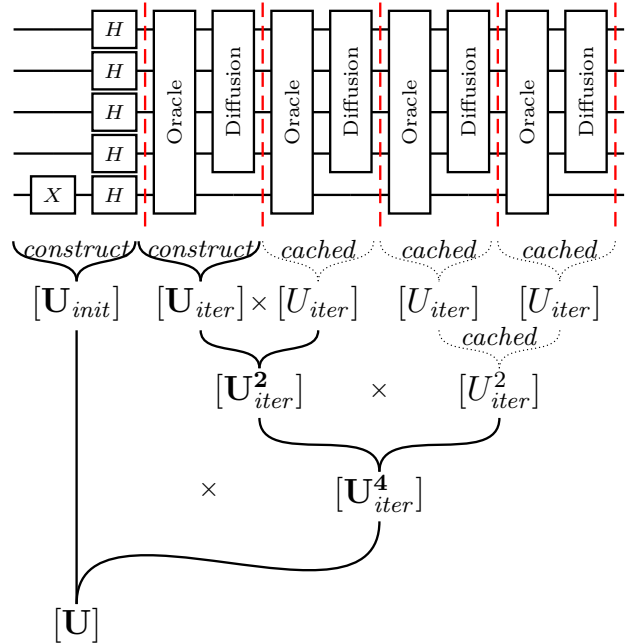
\begin{figure}[t]
	\centering
	\resizebox{0.72\linewidth}{!}{
	\begin{tikzpicture}
	\node[font=\scriptsize] (n0) {
	\begin{quantikz}[column sep=6pt, row sep={0.5cm,between origins}, ampersand replacement=\&]
		\& \qw \&\gate{H}  \slice{} \& \gate[5,disable auto height]{\rotatebox{90}{Oracle}} \& \gate[wires=4,disable auto height]{\rotatebox{90}{Diffusion}} \slice{} \& \gate[wires=5,disable auto height]{\rotatebox{90}{Oracle}} \& \gate[wires=4,disable auto height]{\rotatebox{90}{Diffusion}} \slice{} \& \gate[wires=5,disable auto height]{\rotatebox{90}{Oracle}} \& \gate[wires=4,disable auto height]{\rotatebox{90}{Diffusion}} \slice{} \& \gate[wires=5,disable auto height]{\rotatebox{90}{Oracle}} \& \gate[wires=4,disable auto height]{\rotatebox{90}{Diffusion}}\slice{}  \& \qw \\
		\&\qw \&\gate{H} \& \&  \&  \&  \&  \&  \&  \&  \& \qw \\
		\&\qw \&\gate{H} \&  \&  \&  \&  \&  \&  \&  \&  \& \qw \\
		\&\qw \&\gate{H} \&  \&  \&  \&  \&  \&  \&  \&  \& \qw \\
		\&\gate{X} \&\gate{H} \&  \& \qw \&  \& \qw \&  \& \qw \&  \& \qw  \& \qw
	\end{quantikz}};

	\node (m1) at ($(n0.south west)!0.135!(n0.south east)-(0,6mm)$) {};
	\node (m2) at ($(n0.south west)!0.325!(n0.south east)-(0,6mm)$) {};
	\node (m3) at ($(n0.south west)!0.505!(n0.south east)-(0,6mm)$) {};
	\node (m4) at ($(n0.south west)!0.69!(n0.south east)-(0,6mm)$) {};
	\node (m5) at ($(n0.south west)!0.875!(n0.south east)-(0,6mm)$) {};

	\draw[thick] ($(n0.south west)+(3mm,0)$) to [out=-90, in=90] (m1);
	\draw[thick] ($(n0.south west)!0.225!(n0.south east)$) to [out=-90, in=90] (m1);
	
	\draw[thick] ($(n0.south west)!0.23!(n0.south east)$) to [out=-90, in=90] (m2);
	\draw[thick] ($(n0.south west)!0.415!(n0.south east)$) to [out=-90, in=90] (m2);
	
	\draw[densely dotted] ($(n0.south west)!0.42!(n0.south east)$) to [out=-90, in=90] (m3);
	\draw[densely dotted] ($(n0.south west)!0.59!(n0.south east)$) to [out=-90, in=90] (m3);
	
	\draw[densely dotted] ($(n0.south west)!0.60!(n0.south east)$) to [out=-90, in=90] (m4);
	\draw[densely dotted] ($(n0.south west)!0.77!(n0.south east)$) to [out=-90, in=90] (m4);
	
	\draw[densely dotted] ($(n0.south west)!0.78!(n0.south east)$) to [out=-90, in=90] (m5);
	\draw[densely dotted] ($(n0.south west)!0.96!(n0.south east)$) to [out=-90, in=90] (m5);
	
	\node (t1) at ($(m1)+(0,5.1mm)$) {\emph{construct}};
	\node (t2) at ($(m2)+(0,5.1mm)$) {\emph{construct}};
	\node[] (t3) at ($(m3)+(0,5.1mm)$) {\emph{cached}};
	\node[] (t4) at ($(m4)+(0,5.1mm)$) {\emph{cached}};
	\node[] (t5) at ($(m5)+(0,5.1mm)$) {\emph{cached}};
	
	\node (u1) at ($(m1)-(0,1.5mm)$) {\large $\mathbf{\left[U_{\mathit{init}}\right]}$};
	\node (u2) at ($(m2)-(0,1.5mm)$) {\large $\mathbf{\left[U_{\mathit{iter}}\right]}$};
	\node[] (u3) at ($(m3)-(0,1.5mm)$) {\large $\left[U_{\mathit{iter}}\right]$};
	\node[] (u4) at ($(m4)-(0,1.5mm)$) {\large $\left[U_{\mathit{iter}}\right]$};
	\node[] (u5) at ($(m5)-(0,1.5mm)$) {\large $\left[U_{\mathit{iter}}\right]$};
	
	\node (m6) at ($(u2)!0.5!(u3)-(0,9mm)$) {};
	\node (m7) at ($(u4)!0.5!(u5)-(0,9mm)$) {};
	
	\draw[thick] (u2) to [out=-90, in=90] (m6);
	\draw[thick] (u3) to [out=-90, in=90] (m6);
	
	\draw[densely dotted] (u4) to [out=-90, in=90] (m7);
	\draw[densely dotted] (u5) to [out=-90, in=90] (m7);
	
	\node (t6) at ($(u2)!0.5!(u3)$) {$\mathbf{\times}$};
	\node[] (t7) at ($(m7)+(0,5.mm)$) {\emph{cached}};
	
	\node (u6) at ($(m6)-(0,1.5mm)$) {\large $\mathbf{\left[U_{\mathit{iter}}^2\right]}$};
	\node[] (u7) at ($(m7)-(0,1.5mm)$) {\large $\left[U_{\mathit{iter}}^2\right]$};
	
	\node (m8) at ($(u6)!0.5!(u7)-(0,11mm)$) {};
	
	\draw[thick] (u6) to [out=-90, in=90] (m8);
	\draw[thick] (u7) to [out=-90, in=90] (m8);
	
	\node (t8) at ($(u6)!0.5!(u7)$) {$\mathbf{\times}$};
	
	\node (u8) at ($(m8)-(0,1.5mm)$) {\large $\mathbf{\left[U_{\mathit{iter}}^4\right]}$};
	
	\node (m9) at ($(m1)+(0,-3.9)$) {};
	\node (t9) at ($(m9)+(1.6,1.5)$){$\mathbf{\times}$};
	
	\draw[thick] (u1) to [out=-90, in=90] (m9);
	\draw[thick] (u8) to [out=-90, in=90] (m9);
	
	\node (u9) at ($(m9)-(0,1.4mm)$) {\large $\mathbf{\left[U\right]}$};

	\end{tikzpicture}}
	\caption{Proposed strategy applied to Grover's algorithm}
	\label{fig:proposed_grover}
\end{figure}

\begin{example}\label{ex:proposed_grover}
	In order to illustrate the idea we consider an application of Grover's algorithm~\cite{groverFastQuantumMechanical1996}. 
	The algorithm can be used to search for a specific item in an unstructured set of $N$ items by only querying a given (problem-specific) oracle $\mathcal{O}(\sqrt{N})$ times---a quadratic speed-up over classical methods.
	To this end, it uses $\log(N)+1$ qubits and consists of 
	(1)~a small initialization phase which puts all qubits into an equal superposition (to be represented by~$U_{\mathit{init}}$) and
	(2)~multiple Grover iterations (to be represented by~$U_{\mathit{iter}}$).
	A single Grover iteration consists of querying a given oracle and, afterwards, applying the diffusion operator---effectively increasing the probabilities of states matching the search criterion encoded in the oracle.
	
	Now, consider for example the case $N=16$. This entails $\log(N) + 1 = 5$ qubits and approximately $\sqrt{16} = 4$ Grover iterations. By first constructing the matrices $U_{\mathit{init}}$ and $U_{\mathit{iter}}$ (using the scheme described in Section~\ref{sec:general}) and, then, applying the approach proposed above, an evolution of representations as shown in Fig.~\ref{fig:proposed_grover} results. 
	Here, it can be seen that, at each level, only a single multiplication has to be carried out (while the other multiplications are functionally equivalent and, hence, can be cached/reused). 
	Thus, only three building block multiplications are required in total, while any sequential approach would need four multiplications. Again, this might look negligible for this small example, but has substantial effect once larger instances are considered.
\end{example}

\section{Experimental Evaluations}
\label{sec:results}

In order to experimentally evaluate the proposed approaches, we implemented them on top of the publicly-available JKQ-framework~\cite{willeJKQJKUTools2020} which includes the decision diagram package described in~\cite{zulehnerHowEfficientlyHandle2019} and the state-of-the-art construction approach from~\cite{niemannQMDDsEfficientQuantum2016} as reviewed in Section~\ref{sec:building} and illustrated in Example~\ref{ex:qft_sota}.
The resulting implementation has been integrated into the framework and is available at \url{https://github.com/iic-jku/qfr}. 
Afterwards, we used the resulting implementation to construct representations for the functionality of 
\begin{itemize}
	\item the Quantum Fourier Transform, as a representative for a common building block in quantum algorithms such as Shor's algorithm for integer factorization~\cite{shorPolynomialtimeAlgorithmsPrime1997},
	\item Grover's search algorithm~\cite{groverFastQuantumMechanical1996}, as a representative of an algorithm containing repeated building blocks. 
\end{itemize}
All computations have been performed on a machine with an \mbox{Intel\,i7-6700K} processor and \SI{16}{\gibi\byte} RAM running macOS 11.2.
The obtained results have been split into two parts and are shown in Table~\ref{tab:results}. 
In all tables, $n$ and $m$ denote the number of qubits and the number of gates, respectively. 
Furthermore, the runtime (in CPU seconds) as well as the total memory allocation (in \si{\gibi\byte}) needed to construct the respective representation is listed for

\begin{table}[t]
	\caption{Experimental results}\label{tab:results}
	\sisetup{table-format=4.2, round-mode = places,round-precision = 2, group-minimum-digits = 4, round-integer-to-decimal}
	\captionsetup[subtable]{aboveskip=-0pt,belowskip=-0pt}
	\centering
\begin{subtable}[bt]{0.48\linewidth}
	\centering
	\caption{Results for the QFT}
	\label{tab:resultsqft}
	\small
	\resizebox{0.99\linewidth}{!}{
	\begin{tabular}{rr!{\quad}SS!{\quad}>{\bfseries}SS}\toprule
	\multicolumn{2}{c}{QFT} & \multicolumn{2}{c}{State of the art~\cite{niemannQMDDsEfficientQuantum2016}} & \multicolumn{2}{c}{Prop. scheme~\ref{sec:general}} \\
		\cmidrule(l{1em}r{1em}){1-2}\cmidrule(l{2em}r{2em}){3-4}\cmidrule(l{2em}r{2em}){5-6}
		$n$ & $m$& $t_{\mathit{sota}}$ & $\mathit{mem}_{\mathit{sota}}$ & $t_{\mathit{prop}}$ & $\mathit{mem}_{\mathit{prop}}$\\\midrule
		\begin{filecontents*}{./qft.csv}
12,84,0.027,0.066,0.009,0.065
13,97,0.03,0.07,0.023,0.066
14,112,0.047,0.078,0.02,0.069
15,127,0.154,0.092,0.042,0.076
16,144,0.368,0.094,0.087,0.088
17,161,1.013,0.098,0.251,0.098
18,180,3.789,0.111,1.347,0.12
19,199,10.054,0.163,3.024,0.178
20,220,14.717,0.198,4.079,0.227
21,241,21.014,0.234,7.118,0.291
22,264,27.036,0.271,10.788,0.327
23,287,33.424,0.306,13.49,0.39
24,312,39.829,0.342,14.582,0.391
25,337,46.481,0.379,18.762,0.434
\end{filecontents*}
\csvreader[no head]{./qft.csv}
		{1=\n, 2=\m, 3=\tRef, 4=\mRef, 5=\tProp, 6=\mProp}
		{$\n$ & $\m$ & \tRef & \mRef & \tProp & \mProp \cr}				
		\\[-\normalbaselineskip]\bottomrule		
	\end{tabular}
	}
	\end{subtable}\hfill
\begin{subtable}[bt]{0.51\linewidth}
	\centering
	\caption{Results for Grover's algorithm}
	\label{tab:resultsgrover}
	\small
	\resizebox{\linewidth}{!}{
	\begin{tabular}{rS[table-format=6.0, table-text-alignment = right]!{\quad}S[table-text-alignment = right]S!{\quad}>{\bfseries}SS}\toprule
	\multicolumn{2}{c}{Grover} & \multicolumn{2}{c}{State of the art~\cite{niemannQMDDsEfficientQuantum2016}} & \multicolumn{2}{c}{Prop. scheme~\ref{sec:special}}  \\
		\cmidrule(l{1em}r{1.2em}){1-2}\cmidrule(l{2em}r{2em}){3-4}\cmidrule(l{2em}r{2em}){5-6}
		$n$ & $m$& $t_{\mathit{sota}}$ & $\mathit{mem}_{\mathit{sota}}$ & $t_{\mathit{prop}}$ & $\mathit{mem}_{\mathit{prop}}$ \\\midrule
		\begin{filecontents*}{./grover.csv}
12,1741,0.017,0.066,0.006,0.066
13,2614,0.023,0.067,0.007,0.066
14,3991,0.051,0.072,0.011,0.066
15,6076,0.086,0.076,0.013,0.067
16,9105,0.262,0.092,0.015,0.068
17,13686,0.344,0.083,0.025,0.069
18,20467,3.458,0.095,0.026,0.07
19,30572,3.839,0.098,0.035,0.072
20,45541,26.285,0.099,0.054,0.077
21,67558,68.496,0.102,0.049,0.08
22,100079,361.234,0.108,0.066,0.086
23,147960,\SI{>24}{\hour},{\textemdash},0.082,0.092
24,218425,\SI{>24}{\hour},{\textemdash},0.117,0.099
25,321726,\SI{>24}{\hour},{\textemdash},0.151,0.119
\end{filecontents*}
\csvreader[no head]{./grover.csv}
		{1=\n, 2=\m, 3=\tRef, 4=\mRef, 5=\tProp, 6=\mProp}
		{\n & {\m} & \tRef & \mRef & \tProp & \mProp \cr}					
		\\[-\normalbaselineskip]\bottomrule		
	\end{tabular}
	}
\end{subtable}\\\vspace{1mm}
{\footnotesize $n$: Number of qubits \hspace*{0.3cm} $m$: Number of gates \hspace*{0.3cm} $t$: Runtime in CPU seconds [\si{\second}] \hspace*{0.3cm} $\mathit{mem}$: Total memory allocations [\si{\gibi\byte}]} 
\end{table}

\begin{itemize}
	\item the current state-of-the-art approach~\cite{niemannQMDDsEfficientQuantum2016},
	\item the respective proposed techniques, i.e., the general scheme from Section~\ref{sec:general} in Table~\ref{tab:resultsqft} and the dedicated scheme for repeated structures from Section~\ref{sec:special} in Table~\ref{tab:resultsgrover}.
\end{itemize}

The results for the QFT, which was used as a running example throughout this paper, clearly show that, compared to the current \mbox{state of the art}, the proposed method manages to construct the algorithm's functionality $3.0\times$ faster on average (and up to $4.2\times$ faster).
On the one hand this shows that conducting as many operations as possible on as small as possible intermediate representations indeed pays off.
On the other hand, it confirms the discussion from Section~\ref{sec:general} that although the proposed technique requires to store more representations at the same time, possible redundancies/sharing can explicitly be exploited.

Drastic improvements can be achieved for quantum algorithms containing repeated structures for which the dedicated approach from Section~\ref{sec:special} can be used.
This is confirmed by the numbers provided in Table~\ref{tab:resultsgrover}: Here,  the \mbox{state-of-the-art}
method required \SI{6}{\minute} to construct a representation for the Grover functionality for $n=22$ and failed to construct the functionality at all within \SI{24}{\hour} for larger instances. In contrast, the proposed approach managed to construct the functionality in \emph{all} these cases within fractions of a second. 

In a final series of evaluations, we aimed to compare the proposed techniques to IBM's toolchain Qiskit~\cite{aleksandrowiczQiskitOpensourceFramework2019}, specifically the CPU backend of the Qiskit Aer \emph{UnitarySimulator} in version 0.7.1 which uses a multi-threaded array-based technique for constructing the functionality of a given circuit. 
The results for both the QFT as well as the Grover benchmarks are shown in Table~\ref{tab:resultsqiskit}.
Even for moderately sized instances, we observed runtimes more than two orders of magnitude longer when compared to the technique from~\cite{niemannQMDDsEfficientQuantum2016} or the techniques proposed in this paper.
In addition, IBM's approach requires exponential amount of memory---leading to memory outs when considering more than $15$ qubits while the proposed techniques easily allow to construct the functionality of circuits with more than $20$ qubits.\clearpage

\begin{table}[t]
	\sisetup{table-format=4.2, round-mode = places,round-precision = 2, group-minimum-digits = 4, round-integer-to-decimal}
	\centering
	\caption{Comparison to IBM Qiskit~\cite{aleksandrowiczQiskitOpensourceFramework2019}}
	\label{tab:resultsqiskit}
	\footnotesize
	\resizebox{0.73\linewidth}{!}{
	\begin{tabular}{rS[table-format=6.0, table-text-alignment = right]!{\quad}S[table-text-alignment = right]S!{\quad}rS[round-integer-to-decimal=false, table-format=6.0, table-text-alignment = right]!{\quad}SS}\toprule
	\multicolumn{2}{c}{QFT} & \multicolumn{2}{c}{IBM Qiskit} & \multicolumn{2}{c}{Grover} & \multicolumn{2}{c}{IBM Qiskit} \\
		\cmidrule(l{1em}r{1.2em}){1-2}\cmidrule(l{2em}r{2em}){3-4}\cmidrule(l{2em}r{2em}){5-6}\cmidrule(l{2em}r{2em}){7-8}
		$n$ & $m$& $t$ & $\mathit{mem}$ & $n$ & $m$ & $t$ & $\mathit{mem}$\\\midrule
		\begin{filecontents*}{./qiskit.csv}
12,84,1.8,0.25,10,731,16.7,0.09
13,97,7.9,1.04,11,1112,98.9,0.19
14,112,36,3.92,12,1741,996.38,0.28
15,127,146,15.97,13,2614,11336.69,1.03
16,144,{\textemdash},MemOut,14,3991,\SI{>24}{\hour},3.93
17,161,{\textemdash},MemOut,15,6076,\SI{>24}{\hour},15.94
18,180,{\textemdash},MemOut,16,9105,{\textemdash},MemOut
19,199,{\textemdash},MemOut,17,13686,{\textemdash},MemOut
\end{filecontents*}
\csvreader[no head]{./qiskit.csv}
		{1=\n, 2=\m, 3=\tRef, 4=\mRef, 5=\tProp, 6=\mProp, 7=\tQiskit, 8=\mQiskit}
		{\n & {\m} & \tRef & {\mRef} & \tProp & \mProp & \tQiskit & {\mQiskit} \cr}					
		\\[-\normalbaselineskip]\bottomrule		
	\end{tabular}}\\\vspace{1mm}
	{\footnotesize $n$: Number of qubits \hspace*{0.3cm} $m$: Number of gates \\ $t$: Runtime in CPU seconds [\si{\second}] \hspace*{0.3cm} $\mathit{mem}$: Total memory allocations [\si{\gibi\byte}]} 
\end{table}

\section{Conclusion}
\label{sec:conclusions}
In this work, we addressed the issue of constructing the functional representation of certain building blocks or even entire quantum circuits. Existing approaches for solving this task are severely limited by the rapidly growing size of intermediate representations during the construction.
By conducting as many operations as possible on as small as possible intermediate representations, the solutions proposed in this paper manage to consistently outperform existing approaches---allowing to construct the desired representations several factors faster than with the state of the art. 
Moreover, in case repeating structures are explicitly exploited, the construction of the representation for certain prominent quantum algorithms can be completed within seconds, whereas \mbox{state-of-the-art} approaches fail to construct it within an entire day. 
The comparison with IBM's Qiskit has shown that industrial tools for quantum computing are still in their infancy and would greatly benefit from the integration of existing techniques for efficiently constructing functional representations of quantum circuits---and even more so the techniques proposed in this work.

\vspace{1cm}
\subsubsection*{Acknowledgments}
This project has received funding from the European Research Council (ERC) under the European Union’s Horizon 2020 research and innovation programme (grant agreement No. 101001318).
It has partially been supported by the LIT Secure and Correct Systems Lab funded by the State of Upper Austria as well as by the BMK, BMDW, and the State of Upper Austria in the frame of the COMET program (managed by the FFG).

\printbibliography
\end{document}